\documentclass[fleqn,usenatbib]{mnras}

\usepackage{newtxtext,newtxmath}

\usepackage[T1]{fontenc}

\usepackage{graphicx}	
\usepackage{amsmath}	
\usepackage{booktabs}

\usepackage{hyperref}

\newcommand{\tess}{{\it TESS}}

\newcommand{\gaia}{{\it Gaia}}

\newcommand{\NGTS}{NGTS}

\newcommand{\kms}{km\,s$^{-1}$}

\newcommand{\msun}{\mbox{$\mathrm{M}_{\odot}$}}

\defcitealias{GAIA2018}{G18}
\defcitealias{Kounkel2018}{K18}

\title[NCS V: Rotation in Orion]{NGTS clusters survey -- V: Rotation in the Orion Star-forming Complex}

\author[G. Smith et al.]{
Gareth D. Smith,$^{1}$\thanks{E-mail: gds38@cam.ac.uk}
Edward Gillen,$^{2,1}$\thanks{Winton Fellow}
Simon T. Hodgkin,$^{3}$
Douglas R. Alves,$^{4,2}$ 
David R. Anderson,$^{5,6}$\newauthor
Matthew P. Battley,$^{7}$
Matthew R. Burleigh,$^{8}$
Sarah L. Casewell,$^{8}$
Samuel Gill,$^{5,6}$
Michael R. Goad,$^{8}$ \newauthor
Beth A. Henderson,$^{8}$ 
James S. Jenkins,$^{9,10}$
Alicia Kendall,$^{8}$
Maximiliano Moyano,$^{11}$
Gavin Ramsay,$^{12}$\newauthor
Rosanna H. Tilbrook,$^{8}$
Jose I. Vines,$^{13}$
Richard G. West,$^{5,6}$
Peter J. Wheatley,$^{5,6}$
\\
$^{1}$Astrophysics Group, Cavendish Laboratory, J.J. Thomson Avenue, Cambridge CB3 0HE, UK\\
$^{2}$Astronomy Unit, Queen Mary University of London, Mile End Road, London E1 4NS, UK\\
$^{3}$Institute of Astronomy, Madingley Road, Cambridge, CB3 0HA, UK\\
$^4$Departamento de Astronom\'ia, Universidad de Chile, Casilla 36-D, Santiago, Chile\\
$^{5}$Centre for Exoplanets and Habitability, University of Warwick, Gibbet Hill Road, Coventry CV4 7AL, UK\\
$^{6}$Department of Physics, University of Warwick, Gibbet Hill Road, Coventry CV4 7AL, UK\\
$^7$Observatoire de Genève, Université de Genève, 51 Chemin Pegasi, 1290 Versoix, Switzerland\\
$^{8}$School of Physics and Astronomy, University of Leicester, Leicester, LE1 7RH, UK\\
$^9$N\'ucleo de Astronom\'ia, Facultad de Ingenier\'ia y Ciencias, Universidad Diego Portales, Av. Ej\'ercito 441, Santiago, Chile\\
$^{10}$Centro de Astrof\'isica y Tecnolog\'ias Afines (CATA), Casilla 36-D, Santiago, Chile\\
$^{11}$Instituto de Astronom\'ia, Universidad Cat\'olica del Norte, Angamos 0610, 1270709, Antofagasta, Chile\\
$^{12}$Armagh Observatory and Planetarium, College Hill, Armagh, BT61 9DG, UK\\
$^{13}$Departamento de Astronom\'ia, Universidad de Chile, Casilla 36-D, 7591245, Santiago, Chile\\
}

\date{Accepted XXX. Received YYY; in original form ZZZ}

\pubyear{2022}

\begin{document}
\label{firstpage}
\pagerange{\pageref{firstpage}--\pageref{lastpage}}
\maketitle

\begin{abstract}
We present a study of rotation across 30 square degrees of the Orion Star-forming Complex, following a $\sim$200 d photometric monitoring campaign by the Next Generation Transit Survey (NGTS). From 5749 light curves of Orion members, we report periodic signatures for 2268 objects and analyse rotation period distributions as a function of colour for 1789 stars with spectral types F0--M5. We select candidate members of Orion using \gaia\ data and assign our targets to kinematic sub-groups. We correct for interstellar extinction on a star-by-star basis and determine stellar and cluster ages using magnetic and non-magnetic stellar evolutionary models. Rotation periods generally lie in the range 1--10 d, with only 1.5 per cent of classical T Tauri stars or Class I/II young stellar objects rotating with periods shorter than 1.8 d, compared with 14 per cent of weak-line T Tauri stars or Class III objects. In period--colour space, the rotation period distribution moves towards shorter periods among low-mass (>M2) stars of age 3--6 Myr, compared with those at 1--3 Myr, with no periods longer than 10 d for stars later than M3.5. This could reflect a mass-dependence for the dispersal of circumstellar discs. Finally, we suggest that the turnover (from increasing to decreasing periods) in the period--colour distributions may occur at lower mass for the older-aged population: $\sim$K5 spectral type at 1--3 Myr shifting to $\sim$M1 at 3--6 Myr. 
 
\end{abstract}




\section{Introduction}
\label{sec:intro}
The evolution of a star depends primarily upon its initial mass, metallicity and angular momentum. These properties help to determine the nuclear reaction rates and processes, energy transport modes, pressure support mechanisms and mass loss rates which lead to distinct evolutionary pathways. While high-mass stars without convection zones do not easily spin down, and so spend their lives as rapid rotators, low-mass stars ($\lesssim$ 1.3 \msun) evolve appreciably, affected along the way by both internal and external factors. During formation, stellar angular momentum is linked to the rotation of the parent molecular cloud, gravitational contraction and to interactions between the protostar and accretion disc. Before and during the early main sequence (MS), core--envelope coupling and interior angular momentum transport become increasingly important \citep{Lanzafame2015, Spada2020}, while MS evolution is dominated by spin-down due to mass loss via magnetised stellar winds. 

Rotation is the main driver for the stellar dynamo and magnetic activity, that manifests itself as (among other things) the photospheric star spots from which we can infer rotation rates in time-series photometry. Specifically, it is the longitudinal inhomogeneity of the spots which gives rise to the periodic photometric modulation of interest. 

Observational studies of stellar rotation have identified three main phases in the evolution of angular momentum in low-mass stars: the first few million years of the PMS phase consists of a largely constant surface rotation rate, followed by an abrupt increase towards the zero-age main sequence (ZAMS), and a steady decline on the MS \citep{Gallet2015}. This picture relies on observations of coeval populations of stars in open clusters and interpreting them in a coherent theoretical framework, such that we can attempt to understand the sequence of evolutionary steps and the relative ages at which they take place. Well-populated open clusters are ideal sources, because they contain stars which span a large range of masses, but which are essentially of the same age and composition. For field stars whose properties are otherwise stable (rendering dating by other means such as via isochrones difficult), the empirical MS spin-down can provide a valuable age predictor. It is the basis of gyrochronology \citep{Barnes2003}, potentially opening the door to stellar ages for large datasets (e.g. \citealt{McQuillan2013, McQuillan2014, Angus2015, Davenport2017, Davenport2018, Lu2021}), and a better characterisation of exoplanets and their host stars (e.g. \citealt{Gallet2019, Zhou2021}). Models (e.g. \citealt{Gallet2013, Gallet2015, Amard2016, Amard2019}) have demonstrated that the main PMS and MS evolutionary trends, as seen in observations of rotation in low-mass stars, can be reproduced, although the picture is not complete (e.g. \citealt{Roquette2021, Godoy2021}). 

The census of observed rotation periods for low-mass stars is quite  well-populated at many stages of evolution, from star-forming regions to mature clusters, e.g. at 1 Myr in the Orion Nebula Cluster (ONC) \citep{Herbst2002, Rodriguez2009} and $\rho$Oph \citep{Rebull2018}; 3 Myr in Taurus \citep{Rebull2020} and NGC 2264 \citep{Lamm2004, Venuti2017}; 8 Myr in Upper Sco \citep{Rebull2018}; 13 Myr in h Per \citep{Moraux2013}; 35 Myr in NGC 2547 \citep{Irwin2008}; 110--120 Myr in the Pleiades \citep{Rebull2016A} and Blanco 1 \citep{Gillen2020}; 150 Myr in NGC 2516 \citep{Bouma2021}; 700--800 Myr in Praesepe and the Hyades \citep{Brandt2015a,Brandt2015b}; 1 Gyr in NGC 6811 \citep{Meibom2011}; 3 Gyr in Ruprecht 147 \citep{Gruner2020}; and 4 Gyr in M67 \citep{Esselstein2018}. However, due to variable mass ranges, time coverage, data quality and field contamination between surveys, new contributions and updates remain valuable. This is particularly so in the era of \gaia\ \citep{Gaia2016,Gaia2021}, with its unmatched astrometry and uniform broadband photometry across the sky. 

Rotation periods for low-mass stars in the youngest-observed clusters (1--3 Myr) have often revealed a broad distribution across the observed mass range, with typical periods of 1--10 d (e.g. \citealp{Bouvier2014, Rebull2018, Rebull2020}). The cause of the initial dispersion of rotation rates at ages younger than 1 Myr is not well understood, but it may be linked to star--disc interactions in the embedded protostellar phase, with more massive discs being more efficient at preventing protostars from spinning up \citep{Gallet2013}. 

For young stars, there is significant observational evidence suggesting that, at a given age, stars with a circumstellar disc are, on average, slower rotators than those without a disc (e.g. \citealt{Herbst2002, Rodriguez2009, Affer2013, Serna2021}). Angular momentum evolution models commonly invoke a disc-locking mechanism, caused by magnetic interactions between the young star and its accretion disc, which maintains a constant stellar angular velocity for the lifetime of the inner disc. There are, however, a number of studies which find no significant differences in rotation properties between systems with and without an accretion disc (e.g. \citealt{Stassun1999, Nguyen2009, Blanc2011, Karim2016}). During the early PMS (< 10 Myr), the lowest-mass stars appear to preferentially spin up, leaving a dearth of slow rotators \citep{Bouvier2014, Roquette2021}. Such observations are often correlated with the presence or absence of excess near--mid infrared emission, indicative of an accretion disc \citep{Herbst2002, Rodriguez2009, Affer2013, Venuti2017, Rebull2018}. Hence, they can be explained by different disc-locking timescales, on the understanding that slow rotators are prevented from spinning up due to ongoing star--disc interactions, whilst the discs of fast rotators have been dissipated, allowing the young stars to spin up as they contract towards the ZAMS, which they reach at ages of approximately 22, 33, 66, and 100 Myr for masses of 1.2, 1.0, 0.7, and 0.5 \msun, respectively \citep{Moraux2013}. The mass-dependent disc-locking timescales may be influenced by the local environment, i.e. via external photoevaporation driven by the far-ultraviolet emission of massive stars, which disperses discs around very low-mass stars more quickly than those around higher mass stars \citep{Roquette2021}. At higher masses ($\gtrsim$ 0.3 \msun), bimodal distributions, which may also be a consequence of variations in disc longevity, have been reported, e.g. at $\sim$1 Myr in the ONC \citep{Herbst2002, Rodriguez2009} and $\sim$3 Myr in NGC 2264  \citep{Venuti2017}, as well as at post-accretion ages (> 10 Myr), e.g. at $\sim$13 Myr in h Per \citep{Moraux2013} -- there interpreted as evidence for core--envelope decoupling -- and at $\sim$120 Myr in the Pleiades \citep{Rebull2016A}. 

The Orion Star-forming Complex is one of the largest and most active regions of nearby star formation, comprising numerous well-studied clusters with ages up to $\sim$10 Myr. It is located at an average distance of $\sim$400 pc, toward the Galactic anticentre \citep[hereafter \citetalias{Kounkel2018}]{Kounkel2018}. On the sky, it spans approximately 75 to 90$^{\circ}$ in right ascension and $-10$ to $13^{\circ}$ in declination. The current epoch of star formation is confined to the Orion A and B molecular clouds, home to familiar clusters such as the Orion Nebula Cluster (ONC), NGC 2024, and NGC 2068, where typical ages are 1--3 Myr \citepalias{Kounkel2018}.
\gaia\ astrometry has enabled more extensive and accurate membership lists to be compiled, coupled with a wealth of other photometric and spectroscopic data that have made study of the Orion Complex more tractable.

\citet{Godoy2021} conducted a systematic revision of open cluster sequences based on \gaia\ DR2 and noted that rotation sequences measured from the ground can be as informative for stellar rotation studies as those from space. Furthermore, while observing from space brings many benefits, cutting-edge ground-based facilities can comfortably measure rotation periods from time-series photometry, and can often do so for longer and in more crowded environments than some space-based instruments. This paper presents a study of rotation in Orion using $\sim$200 d of ground-based data from the Next Generation Transit Survey (\NGTS; \citealt{Chazelas2012, Wheatley2018}). The observations were taken as part of the NGTS Clusters Survey (\citealt{Gillen2020, Jackman2020, Smith2021, Moulton2023}).

\section{Observations and target selection}
\begin{figure}
	\includegraphics[width=\columnwidth]{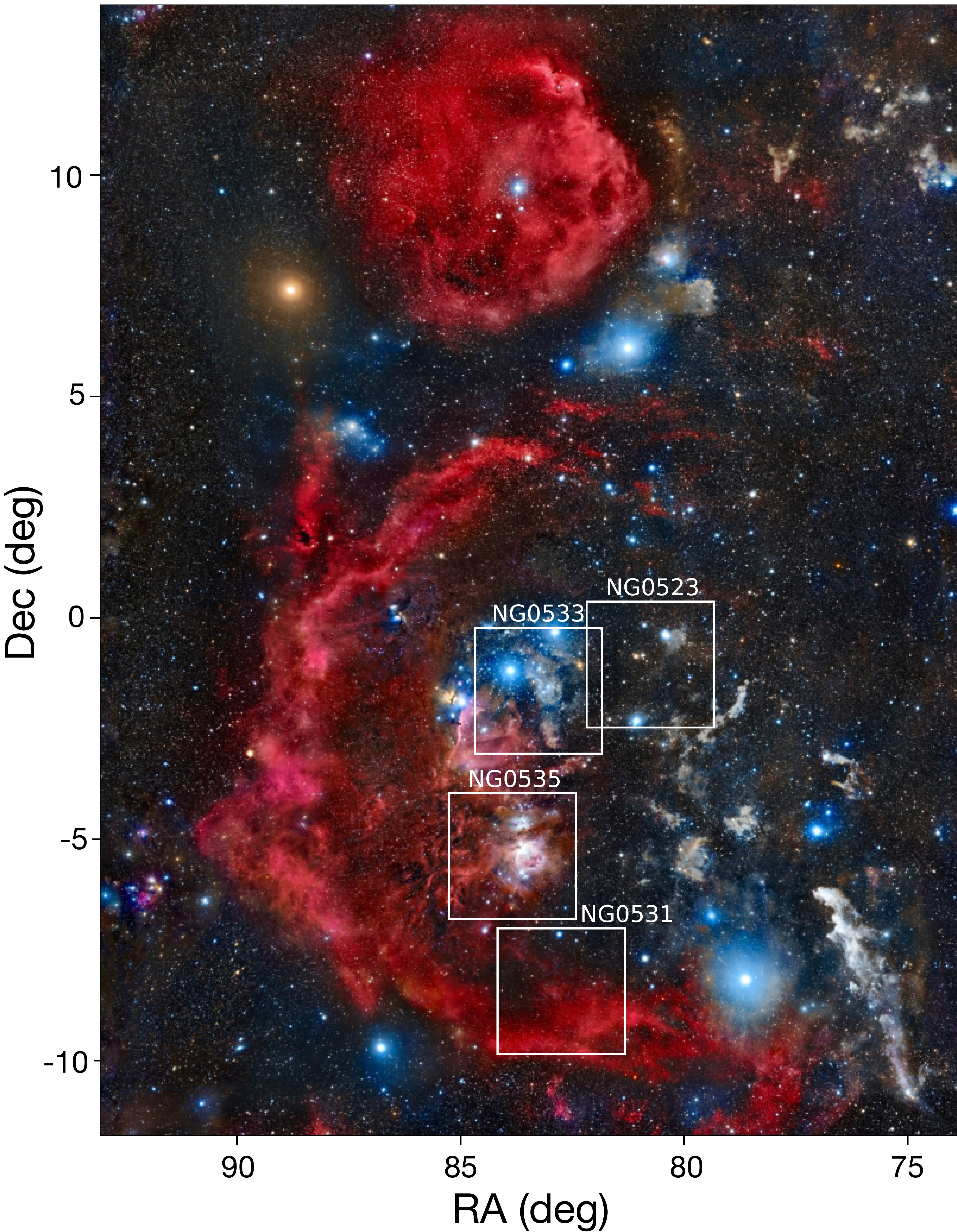}
    \caption{The Orion Star-forming Complex overlaid with rectangles representing the four NGTS fields used in this work. Astrophotograph credit: Rogelio Bernal Andreo (DeepSkyColors.com).} 
    \label{fig:orion}
\end{figure}

To date, NGTS has observed the Orion Complex at four locations, using single cameras with 2.8$^{\circ}$ fields of view (see Figure \ref{fig:orion}), 5-arcsec pixels, point spread functions of below 12 $\mathrm{\mu m}$ (<1 pixel), apertures of radius 3 pixels, a 520--890-nm bandpass, at 13 s cadence, with 10 s exposures. Details of the observations are displayed in Table \ref{tab:obs_details}.
\begin{table}
	\centering
	\caption{NGTS observation details: Full field names, start -- end dates, time baseline in days and number of nights observed.}
    \footnotesize
	\begin{tabular}{llccc}
	Field	&	Dates		& Baseline (d)  &  Nights\\
	\hline
	\hline
    NG0531-0826	& 2015 09 24 -- 2016 03 19 & 177 &  137\\
    NG0535-0523	& 2017 08 17 -- 2018 03 18 & 213 &  156\\
    NG0523-0104	& 2017 08 31 -- 2018 04 08 & 220 & 113\\
    NG0533-0139	& 2020 10 02 -- 2021 04 20 & 200 &  144\\
    \hline
	\end{tabular}
    \label{tab:obs_details}
\end{table}

\subsection{Selection of candidate cluster members}
\label{sec:selection}
\begin{figure}
	\includegraphics[width=\columnwidth]{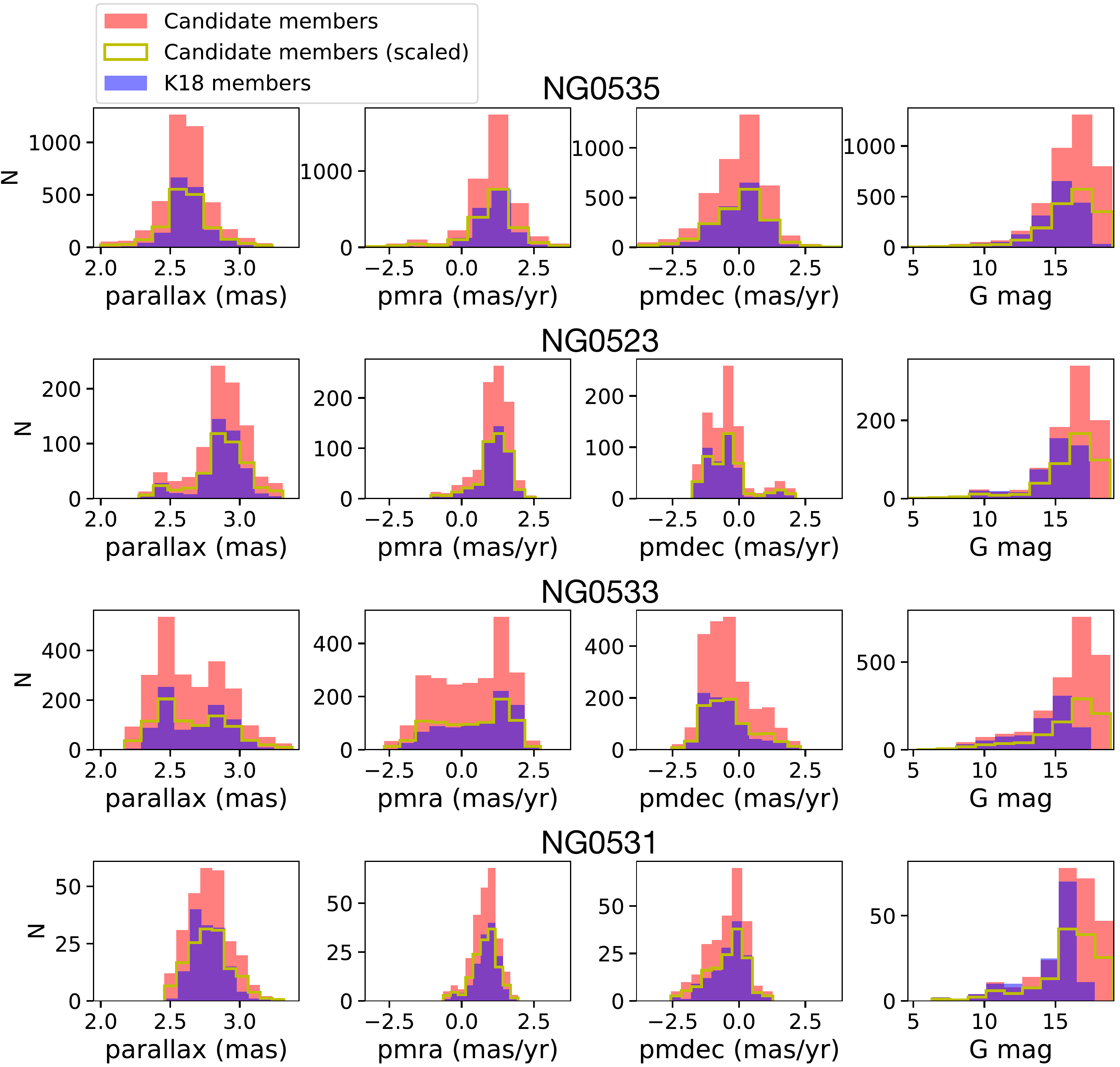}
    \caption{Histograms showing the \gaia\ parallax, proper motion and \gaia\ $G$-mag distributions of the candidate Orion members (red, with scaled version in yellow outline) and the \citetalias{Kounkel2018} Orion members (purple) for the four NGTS fields studied.} 
    \label{fig:candidates_hist}
\end{figure}

\begin{figure}
	\includegraphics[width=\columnwidth]{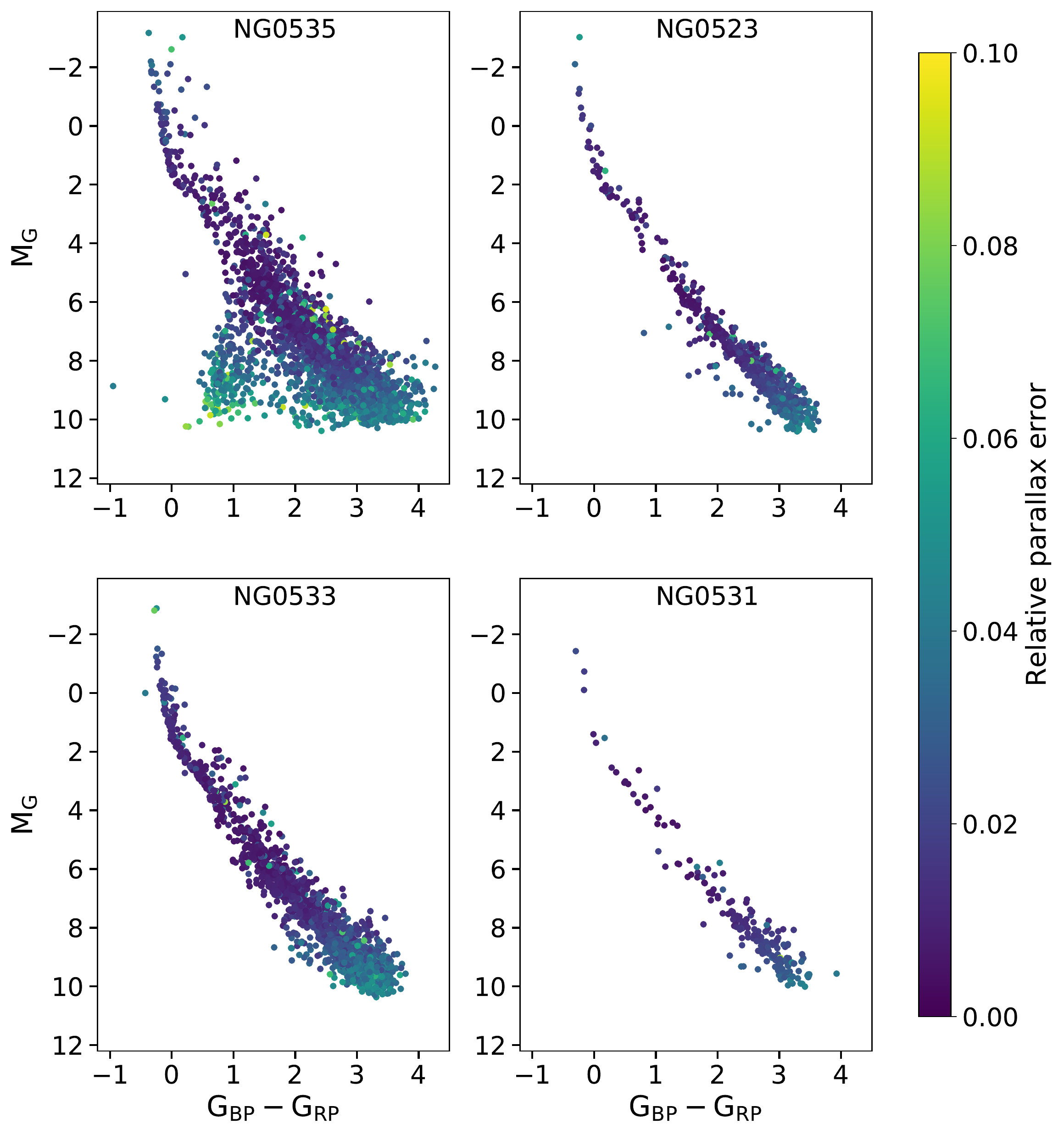}
    \caption{CMDs of the candidate Orion members for the four NGTS fields studied. \gaia\ $G$ magnitudes have been converted to absolute magnitudes via their parallaxes. No extinction correction has been applied at this stage.} 
    \label{fig:candidates_CMDs}
\end{figure}

\citetalias{Kounkel2018} performed a kinematic analysis of the Orion Complex using spectroscopic and astrometric data from APOGEE-2 and \gaia\ DR2. They applied a hierarchical clustering algorithm in five (sometimes six) dimensions to identify distinct groups of young stellar objects (YSOs). Here, we take astrometric and photometric data from the \gaia\ EDR3\footnote{When this work was done, the full DR3 release was not yet available, so it was the EDR3 release from which the \gaia\ measurements were extracted. We note, however, that the astrometry and photometry used are identical between DR3 and EDR3.} data release and create a new candidate membership list, using the \citetalias{Kounkel2018} members\footnote{\citetalias{Kounkel2018} members being those stars assigned to a named group in their paper.} to set bounds on the astrometric parameters of potential members in each field. We performed an EDR3--DR2 crossmatch on the \citetalias{Kounkel2018} members using the gaiaedr3.dr2\_neighbourhood query tool (https://gaia.aip.de/metadata/gaiaedr3/dr2\_neighbourhood/), taking the smallest angular separation match if multiple EDR3 sources matched a DR2 ID. \citetalias{Kounkel2018} objects were discarded if parallax ($\pi$), proper motion ($\mu_{\alpha}$, $\mu_{\delta}$), or photometric data was absent, or if the astrometric precision was such that $\sigma_{\pi}/\pi > 0.1$, $\sigma_{\mu_{\alpha}} > 0.2~\textrm{mas yr}^{-1}$, or $\sigma_{\mu_{\delta}} > 0.2~\textrm{mas yr}^{-1}$ was satisfied. After clipping single outliers in three of the four fields, bounds for the new candidates were set to the \citetalias{Kounkel2018} members' minimum and maximum values of $\pi$, $\mu_{\alpha}$ and $\mu_{\delta}$ for each field, with the exception of field NG0535, which contained a large number of outliers in parallax; here, the bounds were set to the mean $\pm4$ standard deviations. All EDR3 sources were then extracted, subject to the \citetalias{Kounkel2018}-based bounds on parallax and proper motion, and $\sigma_{\pi}/\pi < 0.1$, $\sigma_{\mu_{\alpha}} < 0.2~\textrm{mas yr}^{-1}$, and $\sigma_{\mu_{\delta}} < 0.2~\textrm{mas yr}^{-1}$ as requirements on precision. The EDR3 parallaxes were corrected for the zero-point bias using the expression given in \citet{Lindegren2021}, and the $G$-band magnitudes for sources with six-parameter astrometric solutions were corrected in accordance with \citet{Riello2021}. The distributions of the resulting EDR3 candidates and the \citetalias{Kounkel2018} members are shown in Figure \ref{fig:candidates_hist}, and colour--magnitude diagrams (CMDs) are shown in Figure \ref{fig:candidates_CMDs}. 

The approach of using astrometric cuts based upon the \citetalias{Kounkel2018} members' properties leads to naturally similar, but not identical, distributions. If the goal had been replication of the \citetalias{Kounkel2018} distributions, their clustering algorithm would need to be applied, along with radial velocity data from APOGEE. A significant difference also lies in the use of \gaia\ EDR3 vs DR2, where the former is an expanded catalogue, with a level of precision that brings more objects within the boundaries set; this is particularly evident when comparing the $G$-magnitude histograms of Figure \ref{fig:candidates_hist}. The priority in this work was, rather, to try and capture as many members as possible, maximising the yield of YSO rotation periods. The elevated false positive fraction ought to be mitigated somewhat in the periodic sample, following inspection of light curves and the identification of those objects with rotational modulation patterns characteristic of young stars.

\section{Period detection pipeline}
In this section, we explain the processing of light curves, identify spurious signals, distinguish rotation signatures from other variability, and make a comparison to literature measurements.
\subsection{Light curve pre-processing}
Data points flagged by the NGTS pipeline (e.g. from pixel saturation, blooming spikes, cosmic rays, laser crossing events) and 7-sigma outliers were masked (light curves >80 per cent masked were removed). The light curves were binned in time to 20 min, and those with a single gap greater than half the baseline of the observations, or, with three or more gaps greater than 30 d, were dropped.   

\subsection{Periodic signals}
\label{sub:periodic_signals}
Period measurements were made by calculating the Lomb-Scargle periodogram \citep{Lomb1976, Scargle1982} for each candidate member using the \textsc{astropy} \citep{astropy2013} package, over a search grid covering frequencies 0.001 to 24 $\textrm{d}^{-1}$. Periods corresponding to the highest peaks in the periodograms were taken as provisional rotation periods.\footnote{False-alarm probabilities associated with the highest peaks were all effectively zero, with an extreme outlier maximum value of $10^{-6}$ and a median of $10^{-190}$.}

\begin{figure}
	\includegraphics[width=\columnwidth]{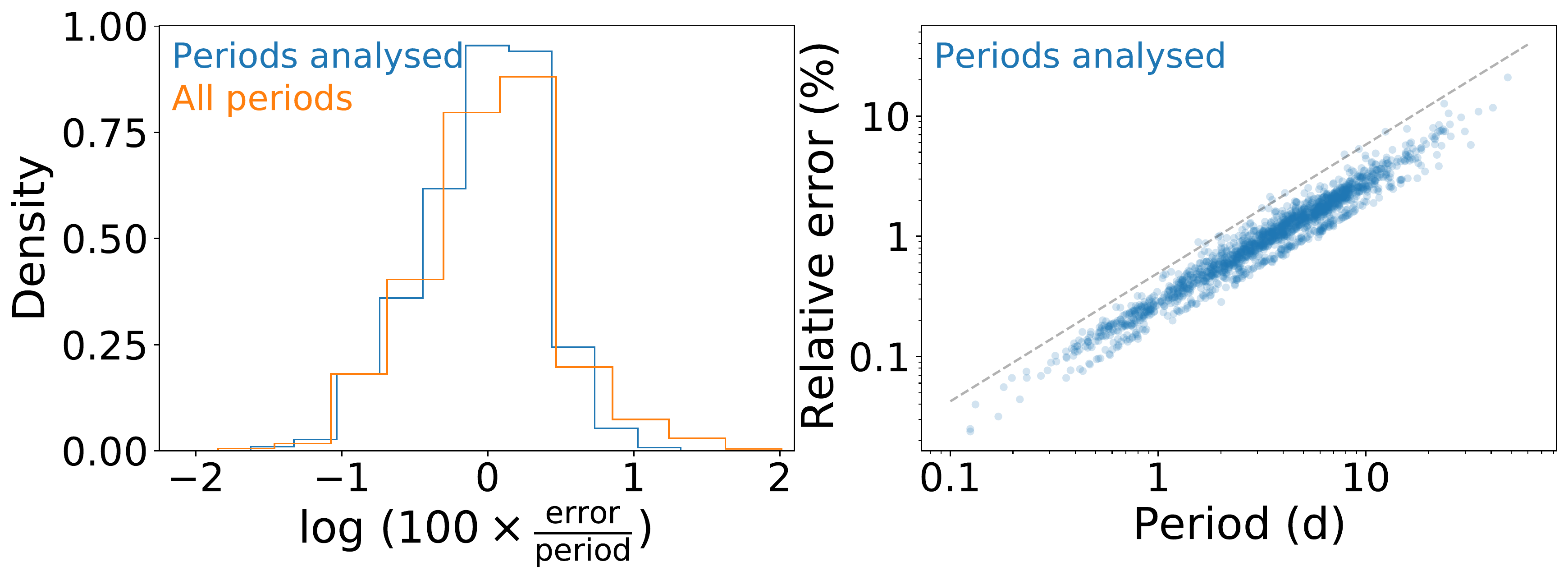}
    \caption{Left: density histogram showing the logarithm of the percentage relative error on the Lomb-Scargle period measurements for the entire periodic sample (orange; $N=2268$) and the periodic sample analysed in period--colour space (blue; $N=1789$; see section \ref{sub:per_col}). Right: percentage relative error as a function of period. The grey dashed line is a precision boundary used during filtering (section \ref{sub:periodic_signals}). We note that the slightly separated population below the main grouping consists of targets observed in field NG0523.} 
    \label{fig:precision}
\end{figure}

While statistical uncertainties on Lomb-Scargle periodogram measurements do not capture the real uncertainties inherent to the technique, e.g. inaccuracies associated with false peaks and aliases, the effects of long-term trends and spot evolution, we measure the half width at half maximum (HWHM) of the periodogram peaks in frequency space to estimate the precision of our measurements (on the assumption that the correct peak has been selected). Figure \ref{fig:precision} summarises these uncertainties\footnote{Using the average of the upper and lower period uncertainties.} by displaying the relative error as a histogram and as a function of period.

NGTS light curves have been found to sometimes retain the imprint of flux from the moon for fainter targets. A typical moon-affected light curve exhibits periodic dips in flux in phase with the lunar cycle, due to an over-correction of the sky background. In order to detect potential rotation signals present in these light curves, a simple trend removal step was incorporated for objects with  $G$ >14 and an initial period between 27 and 30 d. A Savitzky–Golay (SG) filter was applied to the light curves (phase-folded on the detected moon period), followed by a convolution, with the target light curve being detrended by the result of these two steps (SG filter + convolution). An equivalent detrending was applied to light curves with initial periods greater than half the baseline of the observations.

The NGTS pipeline includes a calculation of the dilution affecting each target, with stars within 7 pixels of the target and brighter than magnitude 16 in the \tess\ band contributing. In this work, targets were dropped from the period analysis if the summed flux of the contaminating stars exceeded the target flux. Additionally, objects separated by less than 20 arcsec (a distance at which the flux contribution of an average source falls close to zero) were dropped if the percentage difference in their periods was below the precision boundary line plotted in the right-hand panel of Figure \ref{fig:precision} (at the relevant period), and the measured amplitudes of the signal could not identify the source\footnote{Out of 10 pairs of stars, the amplitude of the shared 
signal was significantly greater on one of the two stars in five cases, such that the corresponding period was retained. Hence, 15 stars were removed from the periodic sample by this procedure.}. Objects with otherwise suspect periods were provisionally rejected, subject to inspection. The periods in question were (1) those likely to be caused by the diurnal pattern of observations, i.e. the one-day signal and its aliases, (2) those ($G$ > 14) which could be an alias of the lunar period (or half the lunar period), and (3) those ($G$ > 14) of $\sim$half the lunar period (13.5--15 d). A detected period was classified as an alias if it fell within calculated boundaries of the expected alias periods described by
\begin{equation}
P_{\mathrm{obs}} = \Bigl(\frac{1}{P_{\mathrm{true}}} + n \Bigr),
\end{equation}
for $n$ in $\pm[1,2,3,4]$. The boundaries were set as
\begin{equation}
\mathrm{Bounds} = \Bigl({P_{\mathrm{obs}}-2\frac{P_{\mathrm{obs}}}{\mathrm{baseline}}}, {P_{\mathrm{obs}}+2\frac{P_{\mathrm{obs}}}{\mathrm{baseline}}}\Bigr),
\end{equation} which comfortably enclosed the corresponding peaks in histograms of the detected periods. The remaining periods were provisionally accepted, subject to inspection.

\begin{figure*}
	\includegraphics[width=0.95\textwidth]{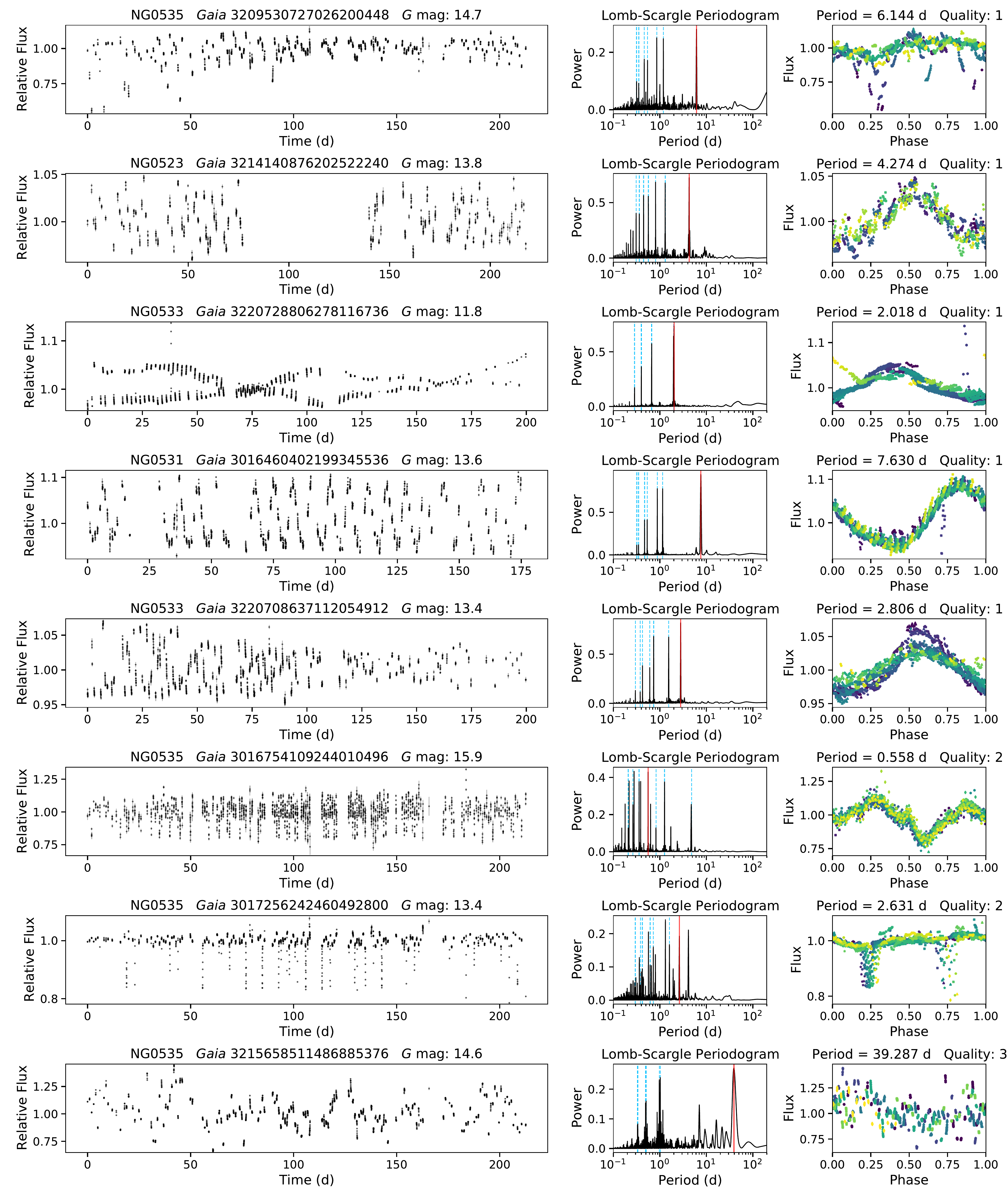}
    \caption{A selection of NGTS light curves and output plots from the periodic detection pipeline. Each object appears in a row, with its light curve (binned to 20 min), followed by a Lomb-Scargle periodogram and light curve phase-folded on the selected period. In each periodogram, a red vertical line locates the periodogram peak of the adopted period, while blue dashed lines locate some of the beat periods resulting from the 1-d sampling. The phase-fold plots cycle through a colourmap with observation time: beginning (blue) to end (yellow). The period and period quality designation (on a scale of 1--4; section \ref{sec:rot_pers}) are shown above. The second and third objects from the bottom are examples where the primary periodogram peak was not selected as the most-likely period. The third-from-bottom star exhibits significant structure in its phase-folded light curve, most likely due to dust enshrouding the system \citep{Stauffer2017, Zhan2019, Gunther2022}, while the system below is an eclipsing binary, where the most-likely period was selected based on the out-of-eclipse variability and differing eclipse depths. The final example is an object where the variability was thought less likely to reflect rotation, with a quality 3 designation given.}
    \label{fig:lightcurves}
\end{figure*}

\subsection{Injection--recovery tests}
In an attempt to evaluate the ability of the period detection pipeline to recover real rotation signals, injection--recovery tests were conducted on each field. For a given stellar magnitude and period, the results take the form of a distribution of threshold amplitudes, above which signal recovery was successful. A percentile score was assigned to each detection in the main sample, based on its amplitude among the test distributions (at the corresponding magnitude and period). We refer the reader to appendix \ref{sec:appendix_inject} and Figure \ref{fig:injection_recovery} for the full details.

\subsection{Rotation periods}
\label{sec:rot_pers}
Pipeline output and light curves (in time and in phase) were inspected, filtered and labelled based on their likelihood of representing stellar rotation. Objects not in one of the three spurious categories previously described were provisionally accepted if either their percentile score from the injection--recovery tests was above 80, or the detection had been labelled as `clean' --  an attempt to identify objects where the detected signal is unique and unambiguous. Following \citet{Xiao2012} and \citet{Covey2016}, the designation is given to objects whose periodogram contains no secondary peaks exceeding 60 per cent of the height of the primary peak, aside from beat periods between the primary peak and the window function (the one-day sampling period). If, upon inspection of the data, a provisionally-accepted signal appeared suspect, that object was then rejected. Objects initially classified as either the one-day signal, an alias of the moon, or half the lunar period, were accepted following inspection in 0.4, 3 and 12 per cent of cases, respectively. This initial stage of inspection left a sample of detections believed to be of astrophysical -- but not necessarily rotational -- origin.

In order to identify the signals which most-likely represent stellar rotation periods, each remaining detection was given a period quality label of 1, 2, 3 or 4. `1' indicates a signal believed to be a clear stellar rotation period (although aliases cannot be ruled out); `2' indicates a signal which is also believed to be stellar rotation, but where the detected signal is relatively weak; `3' indicates a signal which could possibly be stellar rotation, but which could easily be attributed to other forms of variability; and, `4' indicates a signal which, whilst likely to be real, is almost definitely not rotational. Objects in category 3 or 4 tend to have light curves without the typical smooth, starspot-induced modulation patterns for which stellar rotation is a good explanation. They are more stochastic, sometimes displaying signs of accretion bursts or the presence of additional material in the system -- 57 have literature designations of Type I/II YSO or  classical T Tauri star (CTTS), compared with 6 of Type III YSO or weak-line T Tauri star (WTTS). They may also display variations on multiple time-scales, which makes the identification of a rotation signal hard to pinpoint (even if present), particularly for a rigid, single-component model like that used in Lomb-Scargle. The forthcoming analysis of rotation period distributions is restricted to category 1 and 2 objects.

Label 1 was assigned to all objects whose percentile score exceeded 95 and which held the `clean' designation. A second round of inspection identified class 3 and 4 objects, with the remainder being assigned to class 2. Out of 5749 stars with NGTS light curves and 4964 stars for which period measurements were attempted, 2268 periods were retained, with 2179 of those being assigned to class 1 or 2. All period measurements are available in Table \ref{table:data_table}, along with supplementary data on the targets. Figure \ref{fig:lightcurves} shows some example light curves, periodograms and phase folds.

\begin{table*}
    \caption{Data for all candidate Orion members.}
    \begingroup
    \footnotesize
    \begin{tabular}{lll}
    \toprule
     Number &  Column &  Contents   \\
    \midrule
       1 & NGTS ID & NGTS object identification (2102 pipeline run, except NG0533: 2112A pipeline run)  \\
       2 & Field & NGTS observation field  \\ 
       3 & \gaia\ ID & \gaia\ DR3 identification number  \\
       4 & 2MASS ID & 2MASS identification number  \\
       5 & RA & Right ascension (J2000)  \\
       6 & Dec &  Declination (J2000) \\
       7 & Gmag & Stellar magnitude in the \gaia\ $G$ band \\
       8--10 & Period & Rotation period, upper error, lower error\\
       11 & Quality & Designated quality of rotation period (1--4)\\
       12 & Amplitude 90--10 & 90th--10th percentiles of the (relative) stellar flux\\ 
       13 & Literature periods & Rotation periods sourced from the literature\\
       14 & Literature refs & References for literature rotation periods\\
       15--17 & $T_{\mathrm{eff}}$  &  Effective temperature from MCMC posterior distributions (Median, upper error and lower error) \\
       18--20 & $A_{v}$  &  Extinction estimate for the $V$ band from MCMC posterior distributions (Median, upper error and lower error) \\
       21 & N colours & Number of broadband colours used in SED fitting  \\ 
       22 & MCMC success & Whether the MCMC completed successfully: True or False  \\ 
       23 & BP--RP flux excess (corrected) & Corrected BP--RP flux excess as described in \citet{Riello2021}  \\ 
       24 & BP--RP flux excess sigma & N sigma deviation of flux excess from the Stetson and Ivezic standards, as described in \citet{Riello2021}\\ 
       25--29 & $A_{x}$  &  Extinction estimates for the $G, BP, RP, J$ and $H$ bands \\
       30--32 & $(G_{\mathrm{BP}}-G_{\mathrm{RP}})_{0}$ &  \gaia\ BP--RP colour corrected for extinction (Value, upper error and lower error)\\
       33--35 & Luminosity &  Bolometric luminosity as derived from $J$-band or else $G$-band photometry (Value, upper error and lower error) \\
       36 & SED $T_{\mathrm{eff}}$ type  & Source of $T_{\mathrm{eff}}$ constraint in MCMC  \\ 
       37 & TIC-8 $T_{\mathrm{eff}}$ type & TIC-8 source of $T_{\mathrm{eff}}$   \\
       38 & K18 cluster & Assigned sub-cluster in \citetalias{Kounkel2018}  \\ 
       39 & Sub-cluster & Assigned sub-cluster in this work  \\
       40 & Parent cluster & Assigned parent cluster in this work  \\
       41 & Briceno type & T Tauri designations from \citet{Briceno2019}  \\
       42 & Serna type & T Tauri designations from \citet{Serna2021}  \\
       43 & YSO type & YSO designation from \citet{Hernandez2007}, \citet{Megeath2012} or \citet{Marton2016}  \\
       44--48 & HRD age & Value, upper error, lower error, MAD error, fraction of MCMC points within MIST model bounds\\ 
       49--53 & CMD age & Value, upper error, lower error, MAD error, fraction of MCMC points within MIST model bounds\\ 
       54--57 & HRD mass & Value, upper error, lower error, MAD error\\ 
       58--61 & CMD mass & Value, upper error, lower error, MAD error\\ 
       62--66 & HRD cluster age & Value, upper error, lower error, MAD error, number of stars contributing\\ 
       67--71 & CMD cluster age & Value, upper error, lower error, MAD error, number of stars contributing\\ 
       72--76 & HRD sub-cluster age & Value, upper error, lower error, MAD error, number of stars contributing\\ 
       77--81 & CMD sub-cluster age & Value, upper error, lower error, MAD error, number of stars contributing\\ 
       82--96 & Feiden age &Equivalent age data from Feiden magnetic models\\
        \bottomrule
    \noalign{\vskip 2mm}  
    \end{tabular}
    (This table is available in its entirety in machine-readable form) \\
    \endgroup
    \label{table:data_table}
\end{table*}

\subsubsection{NG0523/NG0533 duplicates}
There is a region of overlap between fields NG0523 and NG0533 (see Figure \ref{fig:orion}), resulting in 165 objects with two light curves, 84 of which had period detections in at least one field surviving the above-described filtering. Duplicate objects were removed as follows, based on period detections where available. The most convincing detection was manually selected in seven cases where period estimates disagreed (generally, the differences are attributable to beat periods or harmonics). If an object had a valid period measurement from just one field (10 objects in NG0523 and 11 objects in NG0533), the corresponding data was retained. The data from field NG0533 was preferred in all other cases, due to the large gap in observations for field NG0523. Excluding the seven objects where different Lomb-Scargle peaks were preferred between fields, approximately 100(90) per cent of objects with detections in both fields agree to within 3(1) per cent.

\begin{table}
	\centering
	\caption{Percentages of the Orion candidate members (see section \ref{sec:selection}) with NGTS light curves and with retrieved periods in this work.}
    \footnotesize
	\begin{tabular}{lcc}
	Field	&	In NGTS		& Retrieved periods \\
	\hline
	\hline
    NG0531-0826	&  77\% &  38\%\\
    NG0535-0523	&  82\% &  28\%\\
    NG0523-0104	&  86\% & 41\%\\
    NG0533-0139	&  81\%&  35\%\\
    \hline
	\end{tabular}
    \label{tab:completeness}
\end{table}

\subsubsection{Completeness}
Table \ref{tab:completeness} gives an indication of the completeness of the NGTS sample and the periodic sample as a fraction of the full Orion candidate members list, described in section \ref{sec:selection}. Approximately 80 per cent of the candidate members have NGTS light curves, and we obtain period measurements for $\sim$32 per cent (the accepted periods described above). In the most crowded region -- around the centre of the Trapezium cluster (RA = 83.82, Dec = $-5.39$) -- dilution restricts the number of successfully retrieved periods; although NGTS has light curves for 87 per cent of candidate members in this inner region of the ONC, we retrieve periods for only 15 per cent. As a comparison, observations for the classic \citet{Herbst2002} study of rotation in this same region, were made using 0.24-arcsec pixels, compared with the 5-arcsec pixels of NGTS, which results in a small overlap.

From the candidate members with NGTS light curves, we recover periods for $\sim$40 per cent, whilst from the \citetalias{Kounkel2018} members with light curves (which constitute approximately half of the candidate members with light curves) we recover $\sim$56 per cent\footnote{We note that the fraction of periodic \citetalias{Kounkel2018} members does not increase when we restrict the counts to stars with radial velocity data from APOGEE.}. This difference is partly attributable to the fainter stars incorporated in this work, but may also reflect a smaller false positive fraction of members in the \citetalias{Kounkel2018} sample.

\subsection{Comparison with literature rotation periods}
\begin{figure}
	\includegraphics[width=\columnwidth]{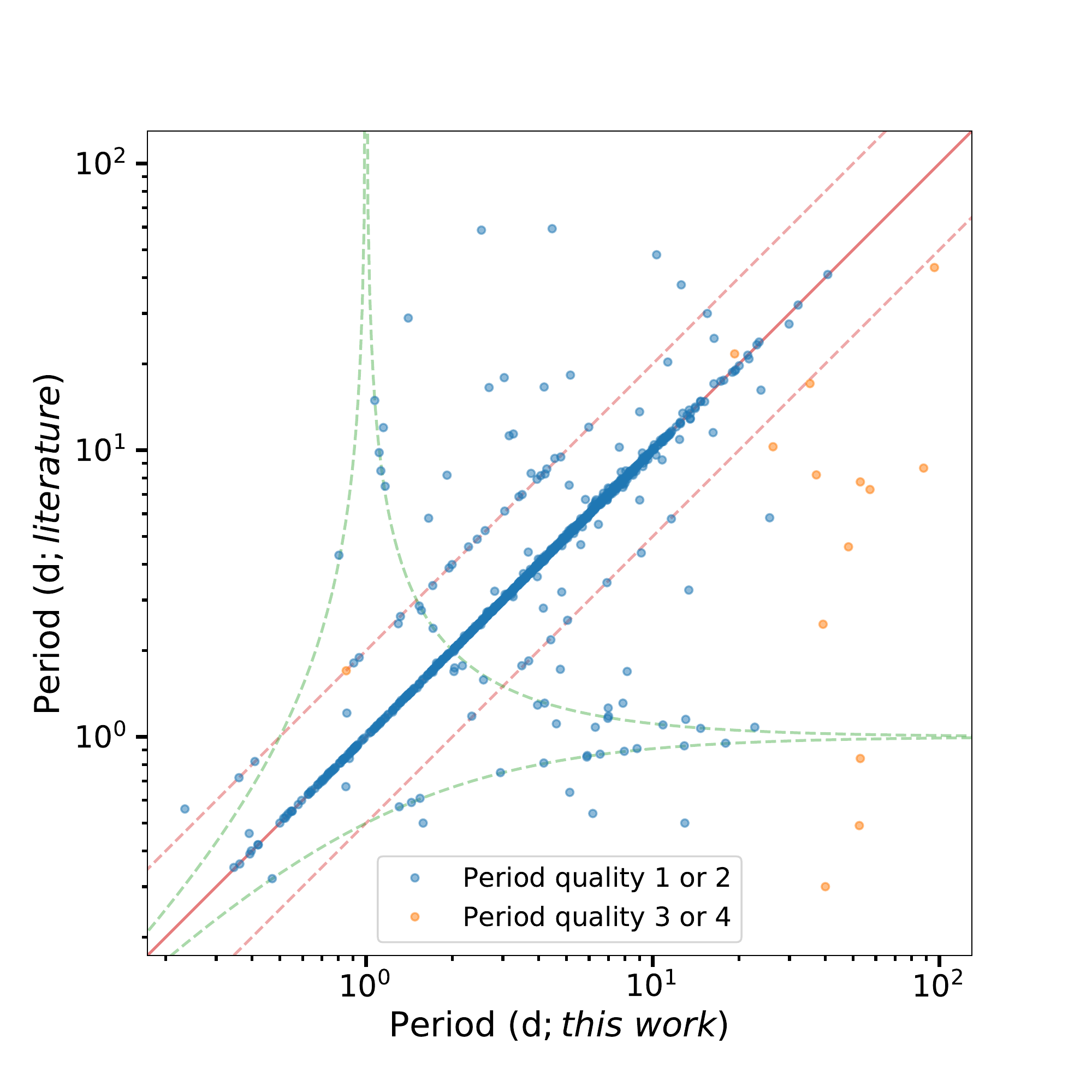}
    \caption{Comparison between literature rotation periods and rotation periods in this work for 957 stars. The solid and two dashed red lines show the 1:1 period match, and the 2:1 and 1:2 harmonics. The dashed green lines show some of the common beat periods inherent to the 1-d sampling of the observations. The period quality categories assigned to periods determined in this work are shown by blue markers (quality 1 or 2) and orange markers (quality 3 or 4). Literature periods were sourced from \citet{Stassun1999}, \citet{Carpenter2001}, \citet{Rebull2001}, \citet{Herbst2002}, \citet{Rebull2006}, \citet{Marilli2007}, \citet{Frasca2009}, \citet{Parihar2009}, \citet{Rodriguez2009}, \citet{Cody2010}, \citet{Morales2011}, \citet{Karim2016}, \citet{Jayasinghe2020} and \citet{Serna2021}.} 
    \label{fig:literature_period_comp}
\end{figure}

We compare our rotation periods to literature values from \citet{Stassun1999}, \citet{Carpenter2001}, \citet{Rebull2001}, \citet{Herbst2002}, \citet{Rebull2006}, \citet{Marilli2007}, \citet{Frasca2009}, \citet{Parihar2009}, \citet{Rodriguez2009}, \citet{Cody2010}, \citet{Morales2011}, \citet{Karim2016}, \citet{Jayasinghe2020} and \citet{Serna2021} (see Figure \ref{fig:literature_period_comp}). Out of the 957 stars in common, we find that 816 (85 per cent) have periods which agree to within 5 per cent. Approximately half of those objects with periods differing by more than 5 per cent are explainable as being either beat periods related to the 1-d sampling of the observations, or as 2:1 or 1:2 harmonics of the periods identified in this work.

\section{Stellar and cluster parameters}
In what follows, we begin by explaining our procedure for estimating interstellar extinction and for obtaining effective temperatures, before assigning stars to kinematic groups and deriving individual and cluster ages.
\subsection{Extinction from broadband photometry}
\label{sec:stellar_params}
We estimated extinction on a star-by-star basis by comparing the observed $G_{\mathrm{BP}}-G_{\mathrm{RP}}$, $G-G_{\mathrm{RP}}$ and $J-H$ colours affected by reddening, with a table of standard colours (SC table hereafter), e.g. $[G_{\mathrm{BP}}-G_{\mathrm{RP}}]_{\mathrm{obs}} - [A_{\mathrm{BP}}-A_{\mathrm{RP}}] = [G_{\mathrm{BP}}-G_{\mathrm{RP}}]_{\mathrm{std}}$, where $A_{\mathrm{BP}}$ and $A_{\mathrm{RP}}$ are the extinctions in the \gaia\ BP and RP photometric bands in this case. The standard colours came from \citet{Luhman2022}. We did not use colours involving WISE or $K$-band photometry, so as to mitigate the worst effects of infrared excess from circumstellar discs. Additionally, we dropped the $J-H$ colour from the fit on occasions when the 2MASS source was matched to multiple \gaia\ objects (a consequence of \textit{Gaia's} higher angular resolution), as determined by the `number\_of\_mates' parameter in the gaiadr3.tmass\_psc\_xsc\_best\_neighbour table from the \gaia\ documentation.

The SC table incorporates the colours and spectral types from Table 4 of \citet{Luhman2022}, and the $T_{\mathrm{eff}}$ values corresponding to spectral types F0--M4 from Table 6 (the empirical < 30 Myr young star table) of \citet{Pecaut2013}. $T_{\mathrm{eff}}$ values for earlier spectral types -- not present in the aforementioned young star table -- were taken from Table 5 (the empirical dwarf table) of \citet{Pecaut2013}\footnote{Updated version available here: \href{https://www.pas.rochester.edu/~emamajek/EEM_dwarf_UBVIJHK_colors_Teff.txt}{Mamajek intrinsic dwarf colours}}. $T_{\mathrm{eff}}$ values for spectral types later than M4 were taken from Table 5 of \citet{Herczeg2014}, following \citet{Fang2017} and \citet{Fang2021}. Linear interpolation in $T_{\mathrm{eff}}$--colour space was applied to obtain a particular intrinsic colour prediction for a given effective temperature. The extinction values were obtained via the reddening law from \citet{Fitzpatrick2019} and synthetically reddened PHOENIX spectra \citep{Husser2013}. A simple Bayesian inference model was employed, with the $T_{\mathrm{eff}}$ and $A_{V}$ posterior parameter space explored using the Markov chain Monte Carlo (MCMC) method implemented in \textsc{Emcee} \citep{Foreman-Mackey2013}.

\subsubsection{Effective temperatures}
Input values and constraints for $T_{\mathrm{eff}}$ were sourced from literature spectral types, from the APOGEE Net pipeline of \citet{Sprague2022}, and from the \tess\ input catalogue, TIC-8 \citep{Stassun2019}. The respective proportions in the final sample of periodic members of Orion were 40, 19 and 39 per cent, with the remaining 2 per cent of objects being fit without constraints on $T_{\mathrm{eff}}$. Two linear corrections were applied to bring the APOGEE Net temperatures onto the same scale as those derived from literature spectral types, whilst a single linear correction was applied to the non-spectroscopically-derived TIC-8 temperatures, a correction only applied to field NG0535, which was the only field showing significant discrepancies. Our adopted uncertainties generally increase with stellar mass and (in the case of spectroscopically-derived temperatures) are typically $\sim$250 K for M spectral types, increasing to $\sim$500 K for spectral types K and G, before increasing steeply for spectral types earlier than mid-F. We refer the reader to appendix \ref{sec:appendix_teff} for details concerning the sourcing of effective temperatures, the corrections, and the derivation of uncertainties. 

\subsubsection{MCMC}
\label{sub:mcmc}
With a small number of exceptions, to be explained in section \ref{sub:ext_constraints}, the MCMC runs were initialised with the derived $T_{\mathrm{eff}}$ values and Gaussian priors described in appendix \ref{sec:appendix_teff}, and a uniform prior on the extinction parameter, $A_{V}$, in the range 0--15 (initial values from the 3D dust maps of \citet{Green2019}). 100 `walkers' explored the posterior parameter space for 5000 steps. The first 3000 steps were discarded as `burn-in', and the values corresponding to the 50th, 84th$-$50th and 50th$-$16th percentiles from the marginalized distributions over the remaining 2000 steps constitute the final adopted values and 1-sigma errors for $T_{\mathrm{eff}}$ and $A_{V}$. For each step in the MCMC, the current value of $A_{V}$ was used to redden a PHOENIX spectrum best-matched with the current value of $T_{\mathrm{eff}}$ and a fixed value of log $g$\footnote{Log $g$, not being well constrained by broadband photometry, was fixed. It was taken from APOGEE Net where available (41 per cent of periodic sample), but in other cases the MIST v1.2 \citep{Mist1, Mist2} stellar models were used to predict its value by interpolation to the star's absolute $J$-band magnitude, derived from 2MASS photometry and \gaia\ parallax at the median age of the stars in the field (based on \citetalias{Kounkel2018} HR ages).}. Reddening of the PHOENIX spectra was applied using the \textsc{Python} \href{https://dust-extinction.readthedocs.io/en/stable/}{Dust Extinction} package. The filter response functions were obtained for each photometric band from the Filter Profile Service, and the effective wavelength for each filter ($\lambdaup$) was calculated using the filter transmission ($T_{\lambdaup}$) and the stellar flux ($S_{\lambdaup}$) in the respective bandpass. The reddening law \citep{Fitzpatrick2019} was finally interpolated to the effective wavelengths for each bandpass to give values for $A_{\lambdaup}/A_{V}$. This process was pre-computed for all bandpasses, for all available PHOENIX spectra, for a range of $A_{V}$ in increments of 0.2. with linear interpolation of all parameters to produce a finer grid.

\subsubsection{Extinction constraints}
\label{sub:ext_constraints}
\begin{figure}
	\includegraphics[width=\columnwidth]{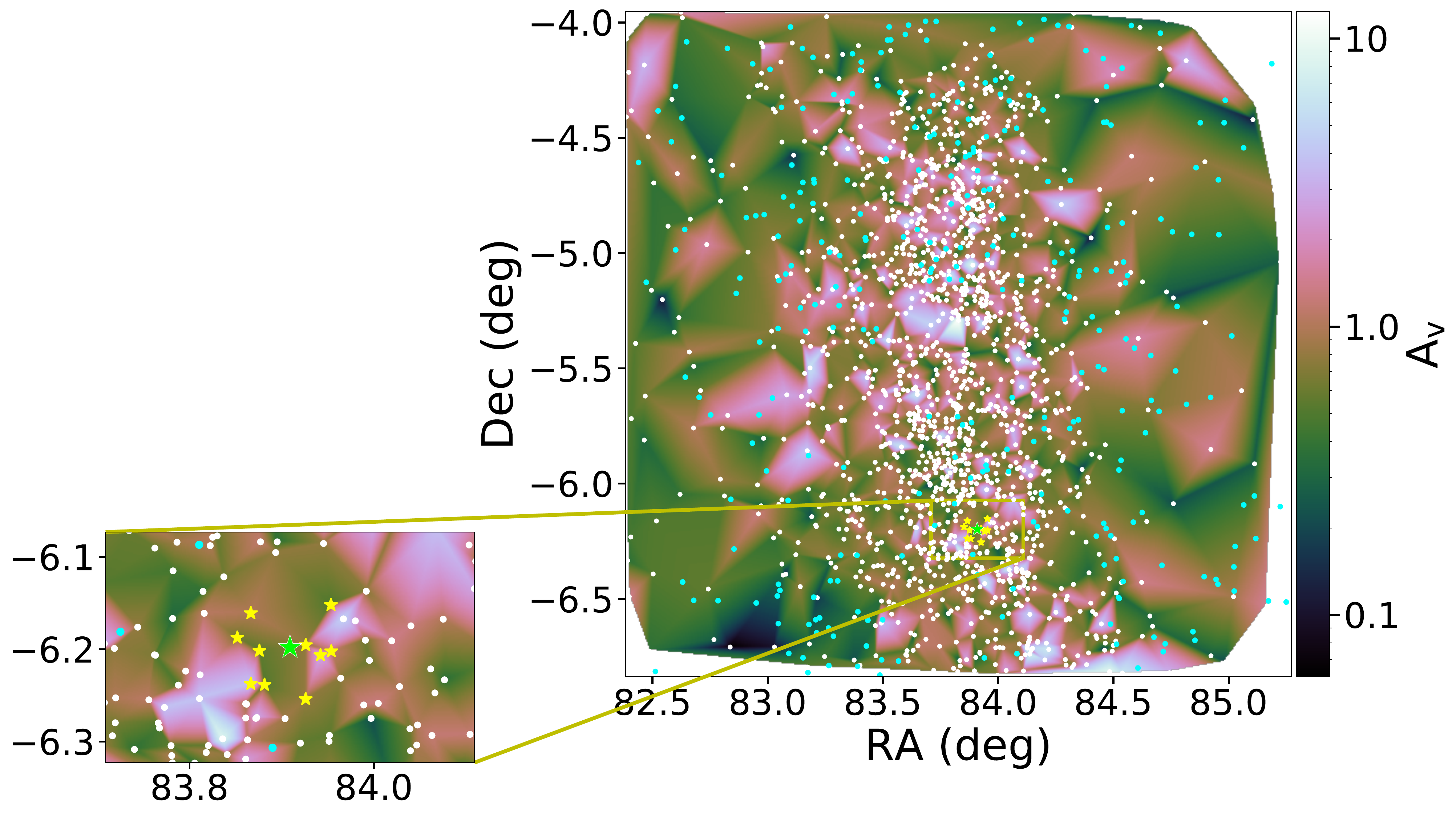}
    \caption{Illustration of $A_{V}$ interpolation in field NG0535. Stars with $A_{V}$ computed through the standard MCMC procedure, i.e. with $T_{\mathrm{eff}}$ constraints from literature spectral types, APOGEE Net, or the spectroscopic TIC-8 sample, are plotted in white. The targets requiring an $A_{V}$ constraint (those with either no available temperature or a non-spectroscopic TIC-8 temperature) are plotted in turquoise. The background colourmap (\textit{cubehelix}; \citealt{DaveGreen2011}) is generated by linear interpolation of the $A_{V}$ of the objects plotted in white. An example target, plotted with a green star, is shown with the 10 nearest objects with measured $A_{V}$ coloured yellow (also shown in zoom-in). The standard deviation of those 10 objects' $A_{V}$ values is used to estimate an uncertainty on the target's interpolated $A_{V}$.}
    \label{fig:av_interp}
\end{figure}

For the objects without any temperature constraint (2 per cent of the sample with rotation periods, and only 0.3 per cent when considering the best-quality sample used for much of the forthcoming analysis), a Gaussian prior was placed on $A_{V}$. For objects in fields NG0523, NG0533 and NG0531, we used the 16th, 50th and 84th percentiles of the reddening predictions from the 3D dust maps of \citet{Green2019}, converting to $A_{V}$ assuming $R_{V}=3.1$ and using the coefficient from Table 6 of \citet{Schlafly2011}. For stars in the ONC-centred field, NG0535, dust map predictions appear to substantially over-predict the extinction, when compared with spectroscopic estimates from \citetalias{Kounkel2018}\footnote{We note that the use of similar dust maps \citep{Green2018} in the dereddening procedure applied in the derivation of photometry-based effective temperatures in the TIC-8 catalogue, could explain the systematically high values when compared with spectroscopically-derived temperatures for field NG0535 (see Figure \ref{fig:TIC_teff_refs}).}. In these cases, the approach adopted was to take the $A_{V}$ values computed from the targets with $T_{\mathrm{eff}}$ constraints from literature spectral types, APOGEE Net, or the spectroscopic TIC-8 sample, and to interpolate in RA--Dec space to the position of each star requiring a constraint on $A_{V}$, i.e. to targets with no available temperature. In addition, due to very large $T_{\mathrm{eff}}$ uncertainties for the objects with non-spectroscopic TIC-8 temperatures (see Figure \ref{fig:tic_sptype_corrections}), a constraint on $A_{V}$ was added to these stars too, in an attempt to break the $T_{\mathrm{eff}}$--$A_{V}$ degeneracy. The Gaussian prior was centred on the interpolated value, with a width equal to the standard deviation of the $A_{V}$ values of the 10 nearest stars to the target. The interpolation was implemented using the \textsc{Scipy} \textsc{Griddata} routine; Figure \ref{fig:av_interp} illustrates the process. Linear interpolation was used, except for the seven stars outside of the interpolation limits; in those cases, the $A_{V}$ value of the nearest neighbour was adopted. To supplement the stars available for interpolation, we included stars not in the membership list, but which still lie within the parallax bounds and meet the requirements on parallax precision. In order to filter out stars with poor quality photometry from the interpolation, we excluded objects by way of a cut on the corrected \gaia\ BP and RP flux excess factor, the cut being made at the 5$\sigma$ level relative to the Stetson and Ivezic standards sample (see equation 18 and section 9.4 of \citealt{Riello2021}).

\subsection{Binary identification}
\label{sec:binaries}
We identify binary and higher-order systems by two methods. Firstly, using the Renormalised Unit Weight Error (RUWE) goodness-of-fit statistic reported in \gaia\ EDR3, which is expected to be around 1.0 for sources where the single-star model provides a good fit to the astrometric observations. We consider objects with a RUWE > 1.4 to be likely binary or higher-order multiple star systems (see e.g. \citealt{Stassun2021}). Secondly, we draw on the spectroscopic analysis of \citet{Kounkel2019} -- a study of high-resolution APOGEE spectra of nearby star-forming regions, searching for double-lined spectroscopic binaries (SB2s). From their catalogue, we take binary candidates to be objects where multiple components were identified in the cross-correlation functions (CCFs), and also objects labelled as `inconclusive SB2/Spotted star pair'. The latter group contains stars where the authors are unsure whether the structure in the CCFs is attributable to multiple stellar components or the impact of star spots, i.e. spots can affect the shape of spectral lines, and hence the CCF profile, as the flux deficit they impart moves across the stellar disc, sometimes resembling a spectroscopically unresolved SB2.

\begin{figure*}
	\includegraphics[width=\textwidth]{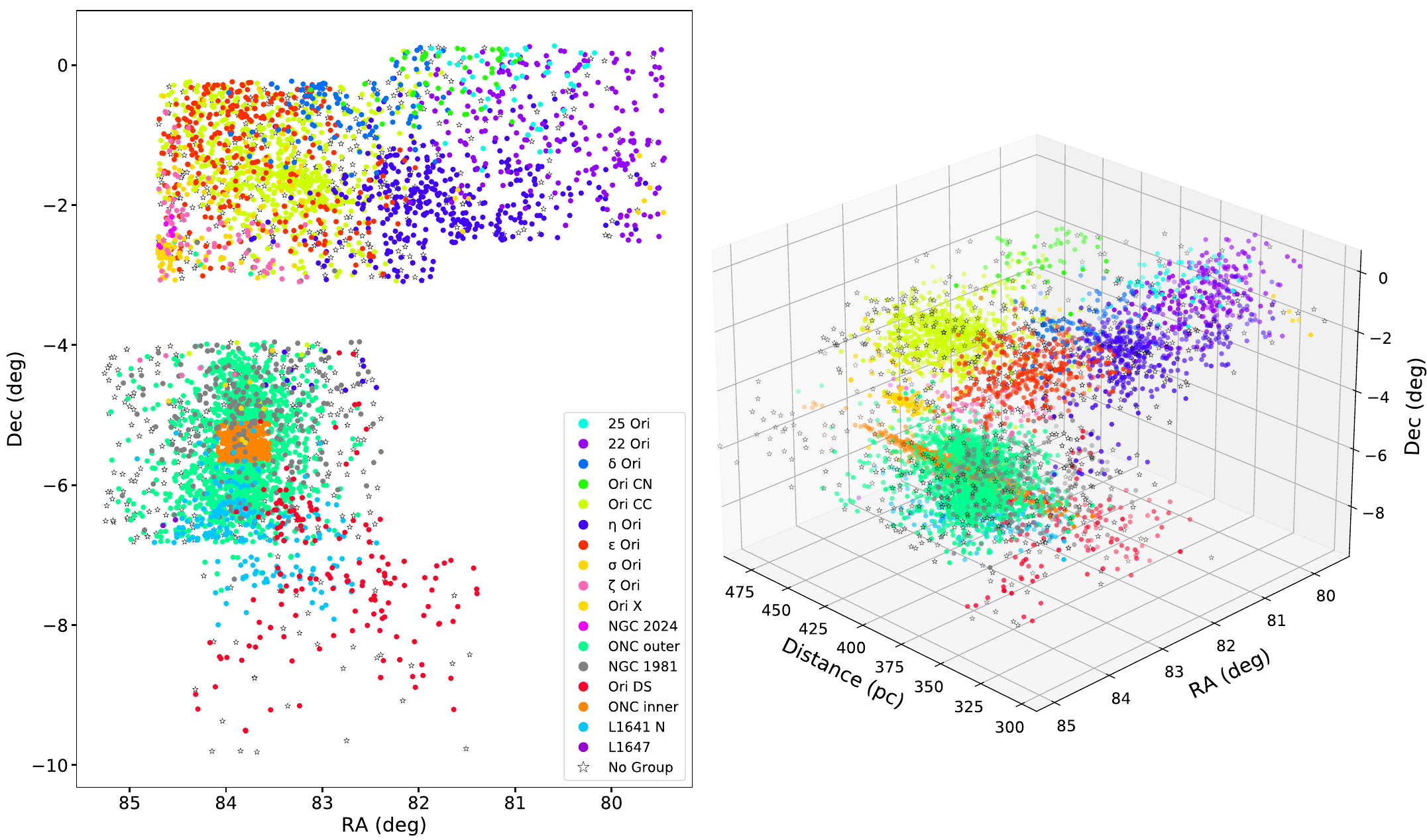}
    \caption{Left: RA--Dec distribution of the candidate Orion member stars with NGTS light curves. Each star is coloured according to its assigned parent cluster. Right: 3D distribution formed by the inclusion of distance. We take distance to be the reciprocal of the \gaia\ parallax, which is reasonable given the median and maximum relative parallax errors of 2 and 9 per cent, respectively. The colour coding is the same as in the left-hand panel.}
    \label{fig:ra_dec_combo}
\end{figure*}

\subsection{Star clusters in Orion}
The Orion Complex is of significant volume and is home to a large number of stellar associations reflecting its star-formation history. These associations, or clusters, represent groups of stars, presumably of very similar age. Hence, we attempted to identify our target stars with their parent cluster. We cross-matched the kinematic groups from \citetalias{Kounkel2018} with \gaia\ EDR3, cutting those outside of the parallax bounds previously described, and those with RUWE > 1.4. Target stars were then assigned to the best-matching group, i.e. the group for which 
\begin{equation}
\chi^{2} = \sum_{i=1}^{5} \Bigl(\frac{x_{i}-\mu_{i}}{\sigma_{i}}\Bigr)^{2}
\end{equation} was minimised, where $x_{i}$, $\sigma_{i}$ and $\mu_{i}$ represent the target value, target error and cluster mean for parameter $i \in \{\mathrm{RA,\, Dec,\, parallax,\, pmRA,\, pmDec}\}$. A further stipulation was that each of the target's astrometric parameters was within the minimum and maximum bounds of the group\footnote{The minimum and maximum values were replaced by the group mean $\pm$ two standard deviations in cases where the latter constituted wider bounds.}. Targets in common with the \citetalias{Kounkel2018} objects were automatically given the \citetalias{Kounkel2018} designation. The \citetalias{Kounkel2018} groups are specified as sub-groups of parent clusters, e.g. `onc-1', `$\sigma$ Ori-3' etc. Collecting these sub-groups into their parent clusters and plotting in 2D and 3D space yields Figure \ref{fig:ra_dec_combo}. The ONC has been split into inner and outer regions, with the inner region being $2000\times2000$ arcsec, centred on the Trapezium cluster\footnote{This matches the location studied by \citet{Herbst2002} in their work on stellar rotation in the ONC.}. 

\subsection{Individual stellar ages}
We derive model-dependent ages (and masses) by linearly interpolating the MIST v1.2 \citep{Mist1, Mist2} stellar evolution models in the HRD (log L vs log $T_{\mathrm{eff}}$) and CMD ($\mathrm{M_{G}}$ vs $\mathrm{M_{BP}-M_{RP}}$). For the HRD, we used the $T_{\mathrm{eff}}$ posterior distributions from the MCMC output, and calculated the stellar luminosities from the absolute extinction-corrected $J$-band magnitudes, using the \gaia\ parallax and the bolometric corrections for PMS stars given by \citet{Pecaut2013}. For stars of spectral type earlier than F, which are absent from the \citet{Pecaut2013} PMS table, the luminosity was calculated based on the \gaia\ $G$-band absolute magnitudes, using the DR3 bolometric correction tool \citep{Creevey2022}. Distributions of luminosities and magnitudes were calculated using the posterior $A_{V}$ distributions, enabling age and mass distributions to be calculated for each target in both the HRD and CMD. We tabulate 50th, 84th$-$50th and 50th$-$16th percentiles of the age and mass distributions, as well as the 1.4826 $\times$ MAD error estimate\footnote{i.e. the median absolute deviation scaled to estimate the standard deviation, assuming normally distributed data.}. In addition, we calculate HRD and CMD quantities based upon the individual median $T_{\mathrm{eff}}$ and $A_{V}$ values from their respective posterior distributions, interpolate to the corresponding single points in the HRD and CMD, and quote these values as our best estimates of age and mass. In the vast majority of cases the values are very similar to taking the median of the age and mass distributions, but when a significant fraction of the points in the HRD or CMD fall outside of the model bounds, the distributions are shifted and are no longer centred on the best estimates of the individual parameters. For that reason, and for reasons of consistency, in what follows we use the age and mass values that are derived from the quoted values of $T_{\mathrm{eff}}$ and $A_{V}$. 

The MIST models were sampled using isochrones between 0.1 and 100 Myr. All post-MS data and parameter space belonging to stars with $T_{\mathrm{eff}}$ > 13 000 K ($M \gtrsim 3.5~\mathrm{M_{\odot}}$; spectral type $\lesssim$ B7.5) was removed. Stars situated in the region corresponding to a younger age than the minimum model age were assigned the minimum age of 0.1 Myr.

\begin{figure}
	\includegraphics[width=\columnwidth]{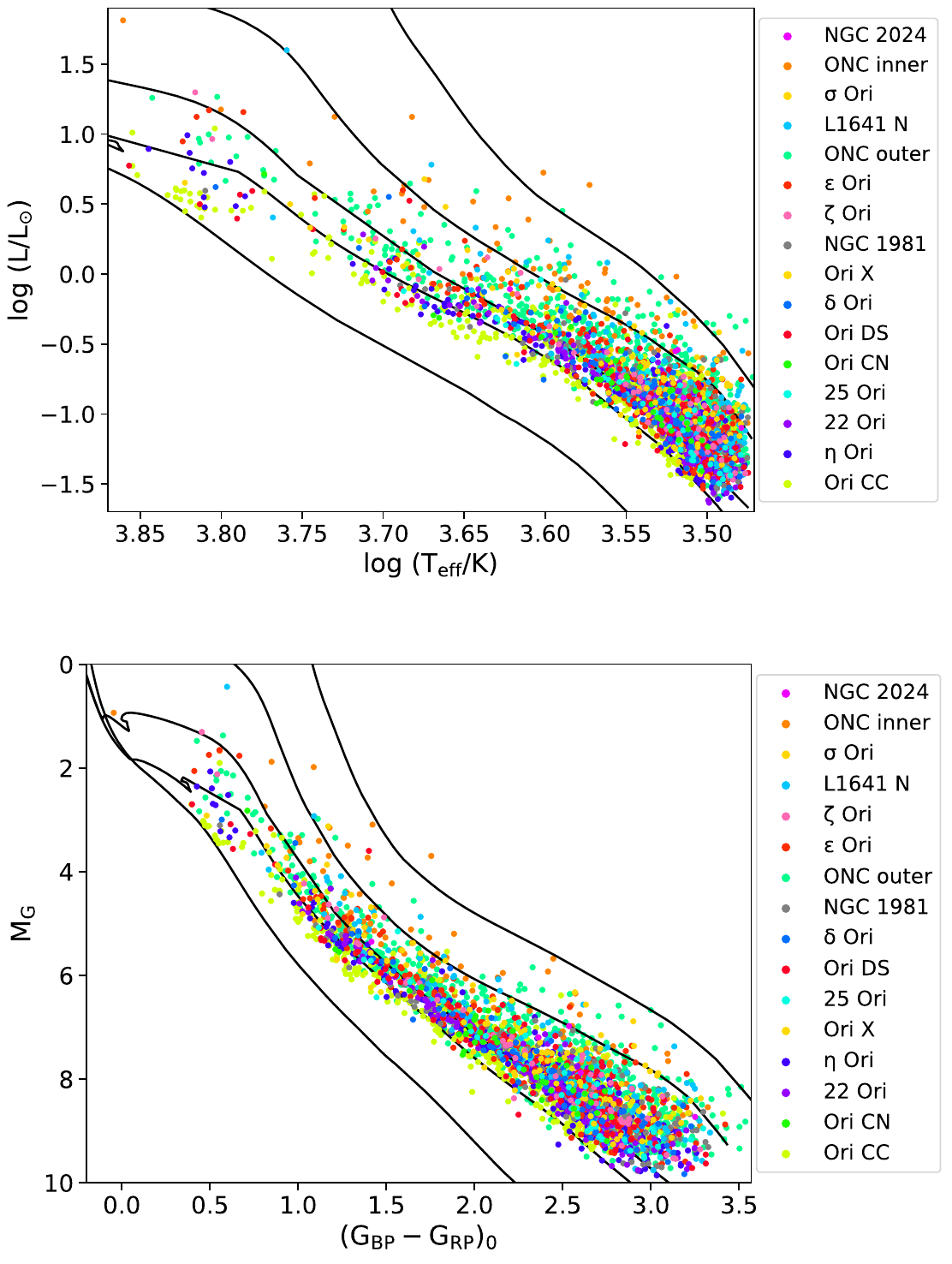}
    \caption{HRD (top) and CMD (bottom) locating the stars used in deriving the cluster ages from the MIST v1.2 \citep{Mist1, Mist2} stellar evolution models (see section \ref{section:cluster_ages}). Points are coloured by their associated parent cluster and the black lines (top right to bottom left) represent the 0.1 Myr, 1 Myr, 5 Myr, 10 Myr and 100 Myr isochrones. A finer grid of model isochrones was used for the interpolation.}
    \label{fig:HRD_CMD}
\end{figure}

\begin{figure*}
    \centering
	\includegraphics[width=0.8\textwidth]{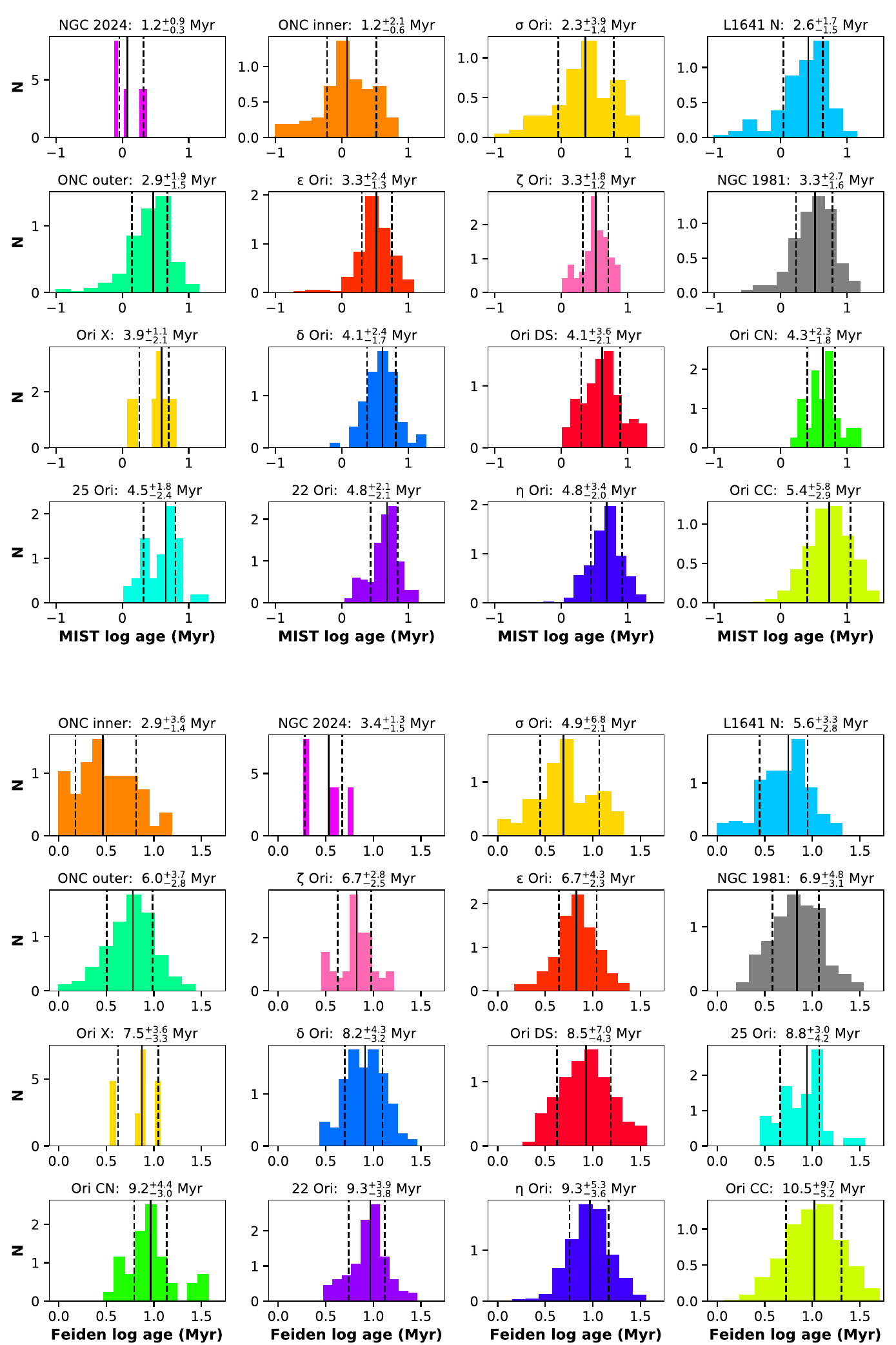}
    \caption{Top: Histograms of the individual stellar ages derived from the HRD and MIST models for the objects used in deriving the cluster ages (see section \ref{section:cluster_ages}). The plots are titled with the cluster name and derived age, and are ordered in increasing age: left-to-right, top-to-bottom. Solid and dashed vertical lines identify the median and 1$\sigma$ uncertainties. Bottom: equivalent histograms using the Feiden magnetic models.}
    \label{fig:individual_age_hists}
\end{figure*}

\begin{figure*}
    \centering
	\includegraphics[width=0.9\textwidth]{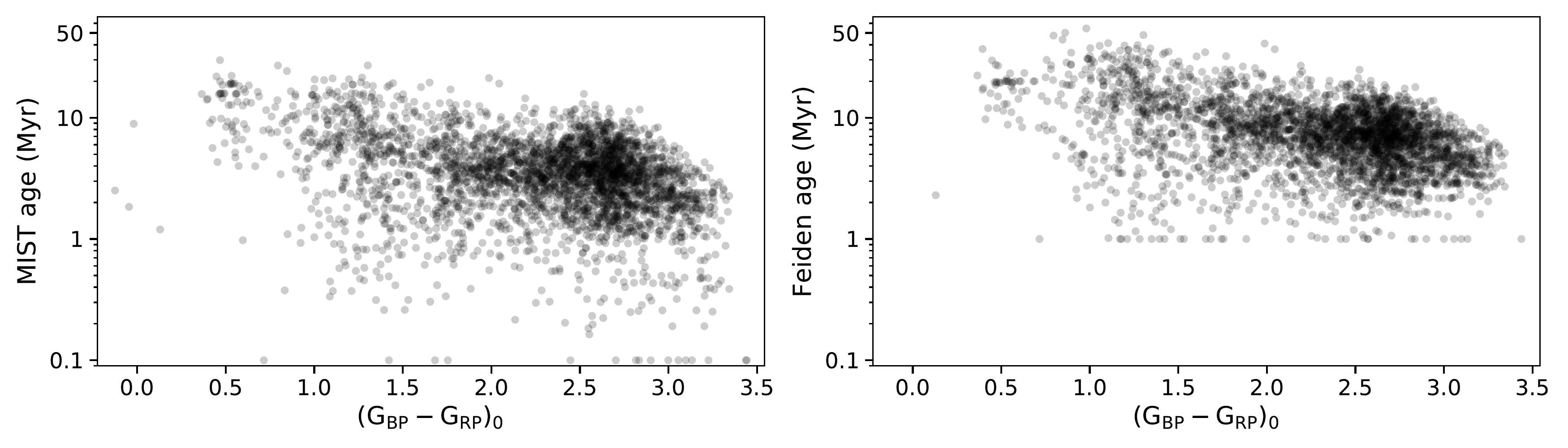}
    \caption{Individual stellar ages derived from the HRD using MIST (left) and Feiden magnetic (right) models vs $(G_{\mathrm{BP}}-G_{\mathrm{RP}})_{0}$ colour for the objects used in deriving cluster ages (see section \ref{section:cluster_ages}).}
    \label{fig:individual_age_colour}
\end{figure*}

\subsection{Cluster ages}
\label{section:cluster_ages}
Cluster ages were taken to be the median age of the corresponding member stars, with upper (84th$-$50th) and lower (50th$-$16th) percentile uncertainties. The member stars were first filtered based on: 
\begin{enumerate}
\item the corrected \gaia\ BP and RP flux excess factor being below the 5$\sigma$ level (as previously described);
\item the fraction of each distribution in the HRD and CMD lying within the model bounds being greater than 0.5\footnote{With the exception of stars whose points fell below the minimum model age.} (which was true and $\sim$1.0 in 91 per cent of cases);
\item the target not being an identified or candidate binary (see section \ref{sec:binaries});
\item the photometry being free from blending issues\footnote{A blended source was taken to be a star with a stellar companion within 1.75 arcsec (limit from \citet{Riello2021}), where the companion was less than three magnitudes fainter in any \gaia\ bandpass. In practice, this cut removed very few additional objects on top of the flux excess filter.};
\item $T_{\mathrm{eff}} < 7280$ K.     
\end{enumerate}
Extreme (7$\sigma$) outliers were also removed (unless the number of cluster members making it through the above cuts was less than five). This process was applied to the parent clusters and to the sub-clusters, for both HRD- and CMD-derived ages. The stars used in the derivation of cluster ages are shown in the HRD and CMD in Figure \ref{fig:HRD_CMD}. HRD-based age estimates are generally found to be younger than those derived from the CMD. In this work we find that the CMD cluster ages are, on average, a factor of 1.2 older than those from the HRD.

Individual HRD stellar ages (for the same stars), as derived from the MIST models, appear as histograms at the top of Figure \ref{fig:individual_age_hists}, and as a function of colour in the left panel of Figure \ref{fig:individual_age_colour}. From Figure \ref{fig:individual_age_colour}, it is apparent that the older stars are, in general, bluer. It is possible that drawbacks to do with the de-reddening method employed, or the more-rapid evolution of higher-mass stars in the HRD, which could make interpolation to model grids more sensitive to any inaccuracies in either observation or theory, could be a factor\footnote{We tested restricting the stars used in deriving cluster ages to those with log $T_{\mathrm{eff}}$ < 3.75, but the differences were minimal.}. However, we note that trends of increasing stellar age with increasing stellar mass in model predictions have been seen many times before, e.g. \citet{Hillenbrand1997, Hillenbrand2008, Herczeg2015, Feiden2016}. It should also be noted that, as a consequence of the location of the NGTS fields, we sample only a small fraction of the members of some clusters. Table \ref{table:data_table} includes our derived stellar and cluster ages.

\citet{Feiden2016} find that their evolutionary models, which incorporate magnetic inhibition of convection, are able to ameliorate the age discrepancy between high- and low-mass stars in the HR diagram for Upper Scorpius. Inhibition of convection produces lower effective temperatures, slowing the contraction rate of young stars. Therefore, stars have a larger radius and a higher luminosity at a given age. The effect is more dramatic for cool, low-mass stars, having relatively little influence on high-mass stars. Hence, a 10 Myr isochrone from the magnetic Feiden models, looks like a 5 Myr isochrone from non-magnetic models for stars with $T_{\mathrm{eff}}$ < 5000 K. To see if the magnetic Feiden models fix the age discrepancy which we observe with our data and the MIST models, we calculate HRD ages with said models, and plot the results in the bottom half of Figure 
\ref{fig:individual_age_hists} and in the right-hand panel of Figure \ref{fig:individual_age_colour}\footnote{We found that Feiden isochrones for ages younger than 1 Myr cross over those of older ages in a way that makes interpolation problematic. Hence, the minimum-age isochrone used was 1 Myr. Any objects in a region of parameter space corresponding to younger ages were assigned an age of 1 Myr. In addition, there are regions of parameter space (high mass, young age) to which the Feiden models do not extend, but the MIST models do. We find $\sim$40 objects in this category, but do not expect this to affect any of our conclusions.}. It is clear, from Figure \ref{fig:individual_age_colour}, that the trend of increasing age with mass remains. It is also evident that, as expected, the ages derived from the magnetic models are older than those from the non-magnetic MIST models. For our purposes, the absolute ages are less important, as we wish to test, primarily, whether there is any noticeable evolution in the period--colour relation. Hence, the sequence of ages is what matters. We see that three pairs of neighbouring clusters (in Figure \ref{fig:individual_age_hists}) switch places, but that the youngest five clusters are identical for both the MIST- and Feiden-derived ages. This means that a division at age $\leq$3 Myr, based upon the MIST models (as is adopted in the subsequent analysis) would be equivalent to a division at age $\leq$6 Myr using the Feiden models.

\begin{figure*}
	\includegraphics[width=\textwidth]{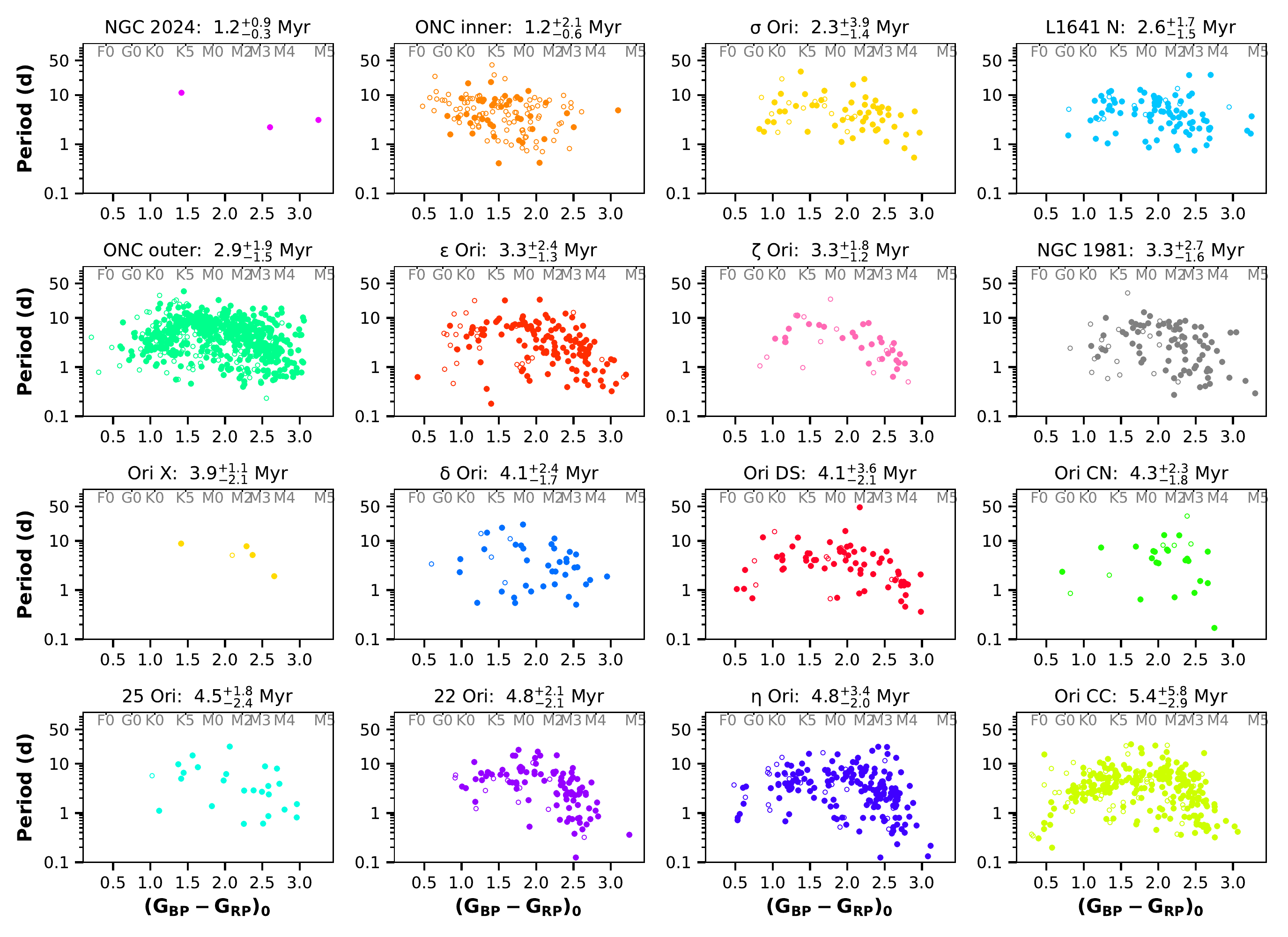}
    \caption{Rotation period vs $(G_{\mathrm{BP}}-G_{\mathrm{RP}})_{0}$ colour for each of the parent clusters, ordered by age (see section \ref{section:cluster_ages}), as derived from the HRD and MIST models. The filled circles represent stars meeting all of the criteria stated in section \ref{section:cluster_ages} and having period quality designation 1 or 2. The open circles are objects which did not meet the stated criteria, but which still have a period quality designation of 1 or 2.}
    \label{fig:period_colour_clusters}
\end{figure*}

We have adopted HRD (rather than CMD) ages in Figures \ref{fig:individual_age_hists}--\ref{fig:period_age} and in the forthcoming analysis for two main reasons: (1) for a clearer comparison between magnetic and non-magnetic evolutionary models (the magnetic models not being available in the colour--magnitude plane); and (2) the age at which to divide samples between old and young being more obvious using the HRD ages. We note, however, that the general age order of clusters is preserved with HRD or CMD ages, bar a few clusters shifting by one or two places, and so the main trends we will highlight in period--colour space (more faster rotators at the blue and red ends and slowest rotators shifting to lower mass for the older populations) are present even if the exact make-up of the young- and old-aged samples changes slightly when adopting CMD instead of HRD ages. Both sets of ages can be found in Table \ref{table:data_table}. With stellar properties, cluster membership and cluster ages determined, we can now investigate the rotation period distribution in Orion. 

\section{Rotation in Orion}
\subsection{Period--colour distribution}
\label{sub:per_col}
The rotation period distribution of each cluster as a function of $(G_{\mathrm{BP}}-G_{\mathrm{RP}})_{0}$ colour is shown in Figure \ref{fig:period_colour_clusters}, ordered by age, as derived from the HRD (see section \ref{section:cluster_ages}). The filled circles represent stars which met all of the criteria stated in section \ref{section:cluster_ages} and have period quality designation 1 or 2. The open circles also have period quality designation 1 or 2, but are objects which did not meet all of the criteria. 

\begin{figure*}
	\includegraphics[width=\textwidth]{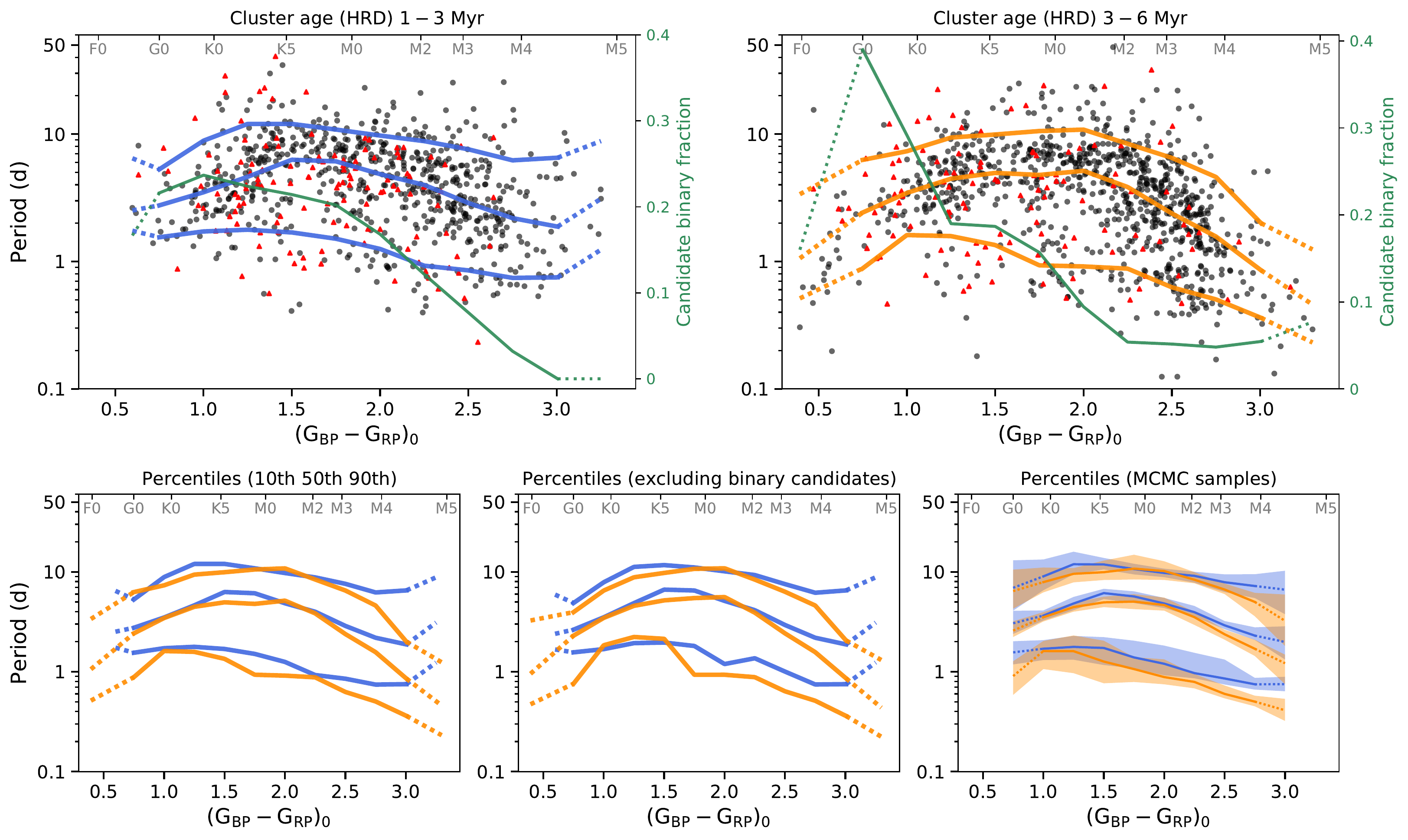}
    \caption{Top: Rotation period vs $(G_{\mathrm{BP}}-G_{\mathrm{RP}})_{0}$ colour for stars belonging to clusters with ages between 1 and 3 Myr (left) and to clusters with ages between 3 and 6 Myr (right). Black circles represent stars with period quality designations 1 or 2 which met the criteria stated in section \ref{section:cluster_ages}. Red triangles are those objects meeting the same criteria except that they are candidate binaries, as per section \ref{sec:binaries}. 10th, 50th and 90th percentiles of the period distributions are overlaid (in blue and orange for the 1--3 and 3--6 Myr ranges, respectively), as are the candidate binary fractions (in green). The bottom-left panel combines the percentiles from the top two plots, while the neighbouring panel shows the percentiles calculated with binary candidates excluded. The bottom-right plot shows the percentile distributions for 1--3 Myr (blue) and 3--6 Myr (orange) populations based on the $\pm 1 \sigma$ distribution of $(G_{\mathrm{BP}}-G_{\mathrm{RP}})_{0}$ from the MCMC samples and the period uncertainties. The median samples are shown by dark lines.}
    \label{fig:period_colour_age}
\end{figure*}

The top row of Figure \ref{fig:period_colour_age} shows the rotation period distribution as a function of $(G_{\mathrm{BP}}-G_{\mathrm{RP}})_{0}$ colour for each of two age groupings: 1--3 Myr on the left and 3--6 Myr on the right. Overlaying the period--colour plots are lines representing the 10th, 50th and 90th percentiles of the rotation period distributions. The black circular markers again represent stars meeting all of the criteria from section \ref{section:cluster_ages}, while the red triangles locate stars which were identified as candidate binaries (see section \ref{sec:binaries}), but which otherwise meet the criteria. The binary candidates  were incorporated into the percentile calculations, which are shown for both ages together in the bottom-left of the figure. Additionally, we show the percentiles with all binary candidates \textit{excluded} in the neighbouring panel. 

In order to assess uncertainties on the percentiles, we generate percentile distributions based on the extinction samples from the MCMC data. We take all $m$ samples from the $\pm 1 \sigma$ region of the corresponding distribution of $(G_{\mathrm{BP}}-G_{\mathrm{RP}})_{0}$ for each star, shuffling the order, giving $n \times m$ samples. We then calculate the 10th, 50th and 90th percentiles of the rotation period distributions for the 1--3 Myr and 3--6 Myr populations $m$ times, i.e. each sample is used once. We incorporate period uncertainties by taking each period to be a random draw from the Gaussian distribution derived by fitting the relevant periodogram peak in frequency space\footnote{Uncertainty on colour due to the uncertainty on extinction is, however, the dominant uncertainty. Typical uncertainties on $(G_{\mathrm{BP}}-G_{\mathrm{RP}})_{0}$ are 0.2--0.4.}. We repeat the process for five separate shuffles of the $(G_{\mathrm{BP}}-G_{\mathrm{RP}})_{0}$ samples and then plot the full extent of the resulting percentile distributions (bottom-right panel of Figure \ref{fig:period_colour_age}).\footnote{All of the percentiles in Figure \ref{fig:period_colour_age} were calculated with a rolling window of width 0.5 in $(G_{\mathrm{BP}}-G_{\mathrm{RP}})_{0}$ colour. Centre-of-bin plotting was applied, except at the extremes of the distributions (shown with dotted lines), where there is left-side-of-bin plotting at the blue end and right-side-of-bin plotting at the red end, with bin widths of 0.25.} 

One of the most prominent differences between the rotation distributions at 1--3 and 3--6 Myr is at the red end, where $(G_{\mathrm{BP}}-G_{\mathrm{RP}})_{0}\gtrsim2.25$ (M2 spectral type). Here, we see a distribution shifted towards shorter rotation periods in the older-age population, with no periods longer than 10 d for $(G_{\mathrm{BP}}-G_{\mathrm{RP}})_{0}>2.65$ (M3.5). The median rotation period at 1--3 Myr decreases from 4 d to 2 d in the range $2.25<(G_{\mathrm{BP}}-G_{\mathrm{RP}})_{0}<3$, but the equivalent decrease at 3--6 Myr is from 4 d to 0.9 d, with additional faster rotators at redder colours. One possible explanation is that the circumstellar discs of very low-mass stars could be more readily dispersed, facilitating an earlier spin-up as they continue their contraction towards the main sequence (e.g. \citealt{Roquette2021}). We also see a population of high-mass, fast rotators in the older-aged group, not present in the younger ensemble.

In order to test the shift towards shorter rotation periods for the older stars, we ran permutation tests comparing the 10th, 50th and 90th percentiles of the young and old populations for $(G_{\mathrm{BP}}-G_{\mathrm{RP}})_{0}$ > 2.25. The tests were run $m$ times: once for each of the $(G_{\mathrm{BP}}-G_{\mathrm{RP}})_{0}$ sample sets described above. 99 per cent of the resulting p values lay below 0.001 and 0.004 for the 10th and 50th percentile tests, respectively, with $\sim$85 per cent of the p values from the 90th percentile tests below 0.05. Hence, these results favour the alternative hypothesis that the rotation periods are longer for the younger population.

Another feature of Figure \ref{fig:period_colour_age} is that the turnover from increasing to decreasing periods is located at lower mass for the older-aged ensemble: $(G_{\mathrm{BP}}-G_{\mathrm{RP}})_{0}\approx2$ (M1 spectral type) at 3--6 Myr, compared with $(G_{\mathrm{BP}}-G_{\mathrm{RP}})_{0}\approx1.5$ (K5) at 1--3 Myr. From the percentile distributions described above, the turnover (as assessed by the 50th percentile) is found at lower mass for the older-aged population 60 per cent of the time, at higher mass 3 per cent of the time, and at approximately the same mass for the remainder. We note that the percentile bin size is similar to the average uncertainty on $(G_{\mathrm{BP}}-G_{\mathrm{RP}})_{0}$, hence there is an issue of resolution. It would be interesting for future rotation studies of young clusters to further investigate this feature.  

Figure \ref{fig:period_colour_age} (top row) also depicts a decreasing fraction of binary candidates moving towards redder colours, which marries with previous findings that bluer, more massive stars are more likely to have companions than redder, less massive ones \citep{Raghavan2010, Duchene2013, Lee2020, Belokurov2020}. However, we also expect binaries to be more difficult to detect in observations of fainter targets, where the signal-to-noise ratio is less favourable. The most prominent difference between the candidate binary fractions of the two age groups is the spike at $(G_{\mathrm{BP}}-G_{\mathrm{RP}})_{0}\lesssim 1$ in the 3--6 Myr sample. However, the impact on the percentiles is small, noticeably affecting only the 10th percentile of the 3--6 Myr group for $1\lesssim (G_{\mathrm{BP}}-G_{\mathrm{RP}})_{0}\lesssim 1.5$. 

\begin{figure}
	\includegraphics[width=\columnwidth]{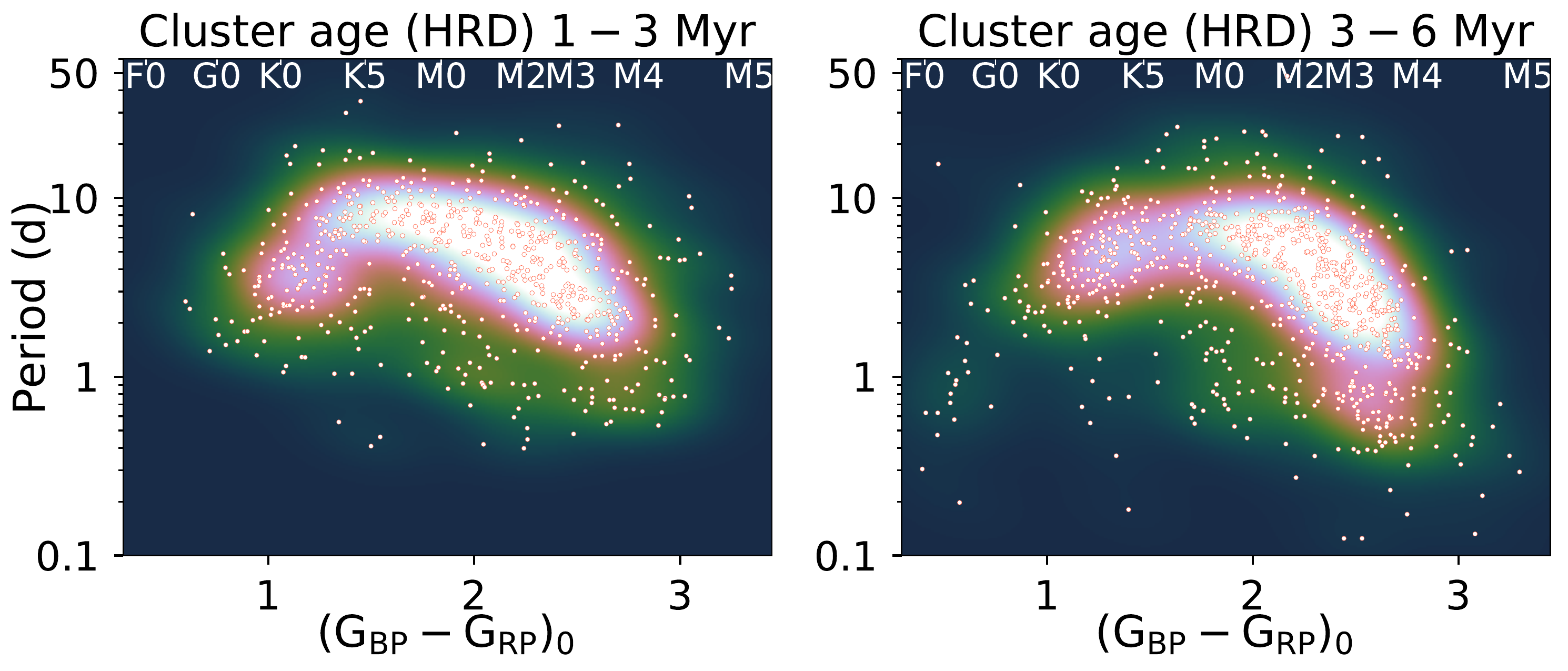}
    \caption{Rotation period vs $(G_{\mathrm{BP}}-G_{\mathrm{RP}})_{0}$ colour overlaying a density map, where brighter shading indicates regions with a greater concentration of points.}
    \label{fig:period_colour_density}
\end{figure}

Figure \ref{fig:period_colour_density} shows the same period--colour distributions (minus the binary candidates), but this time overlaying a density map to more clearly highlight the relative concentration of points across period--colour space. From the density distributions, the steeper slope and later turn-over at the red end (in the older population) are emphasised. We also see the density distribution for the older-aged population extend to shorter periods at the blue and red end.

\subsection{Period--age distribution}
\begin{figure}
	\includegraphics[width=0.9\columnwidth]{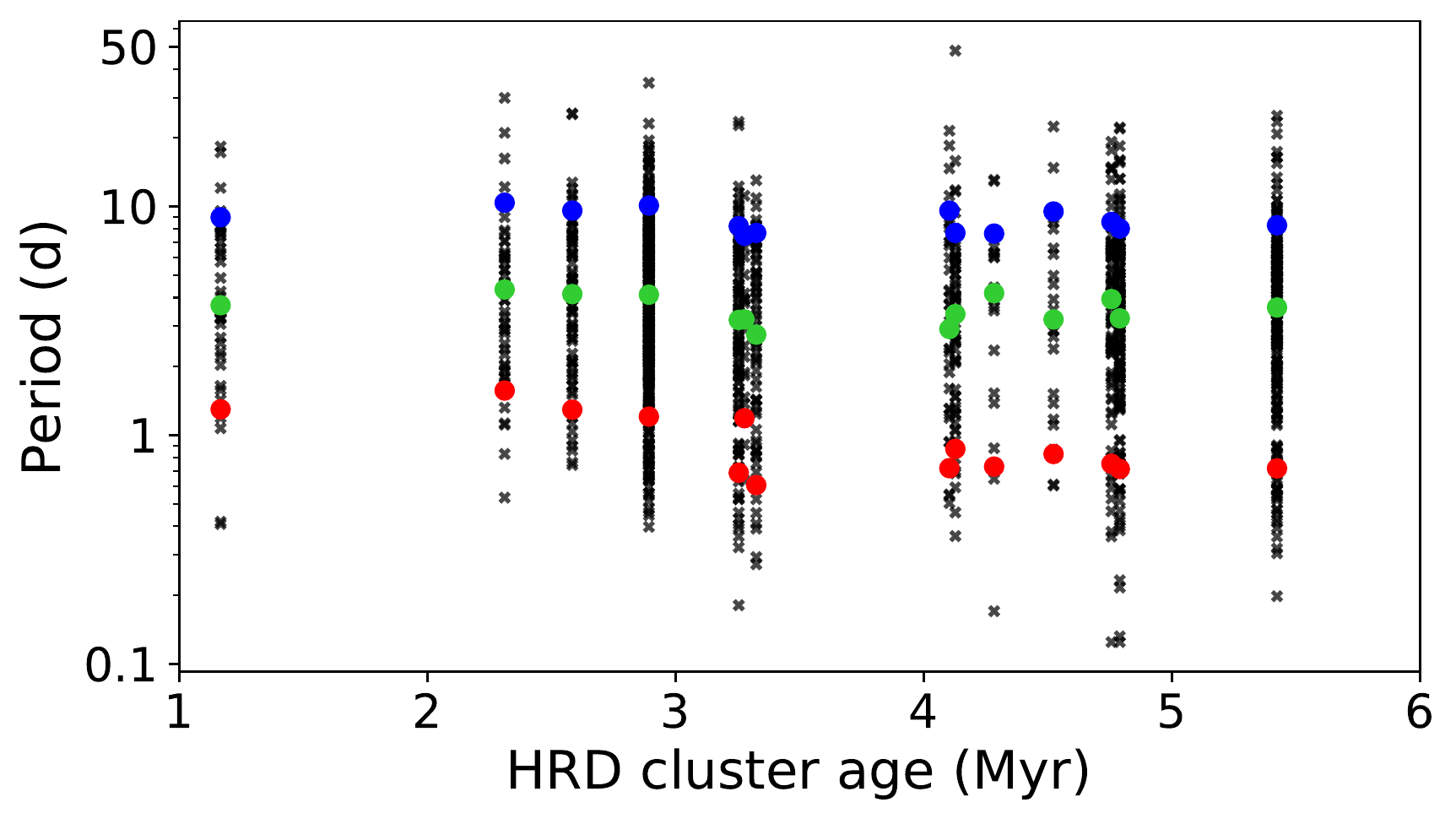}
    \caption{Period vs cluster age for 14 of the 16 parent clusters. NGC 2024 and Ori X are omitted for having very few stars with measured periods in this work. Black crosses locate the periods of the member stars, with 10th, 50th and 90th percentiles shown by red, green and blue circles.}
    \label{fig:period_age}
\end{figure}

Figure \ref{fig:period_age} displays the rotation periods as a function of cluster age\footnote{All stars in a cluster are plotted at a single age (the median value as determined in section \ref{section:cluster_ages}) for clarity.}. Red, green and blue coloured circles mark the 10th, 50th and 90th percentiles of the period distribution for each cluster. The lower envelope of the distributions, as described by the 10th percentiles, transitions to slightly shorter periods after 3 Myr, in line with the observations in period--colour space: $P_{\mathrm{median}} ^{<3\, \mathrm{Myr}}\approx1\, \mathrm{d}$ and $P_{\mathrm{median}} ^{>3\, \mathrm{Myr}}\approx0.7\, \mathrm{d}$. We note that in plotting all cluster members at a single age, the uncertainty and spread in ages is not represented. The figure is nonetheless informative, so long as the sequence of cluster ages is accurate.

\begin{figure}
	\includegraphics[width=\columnwidth]{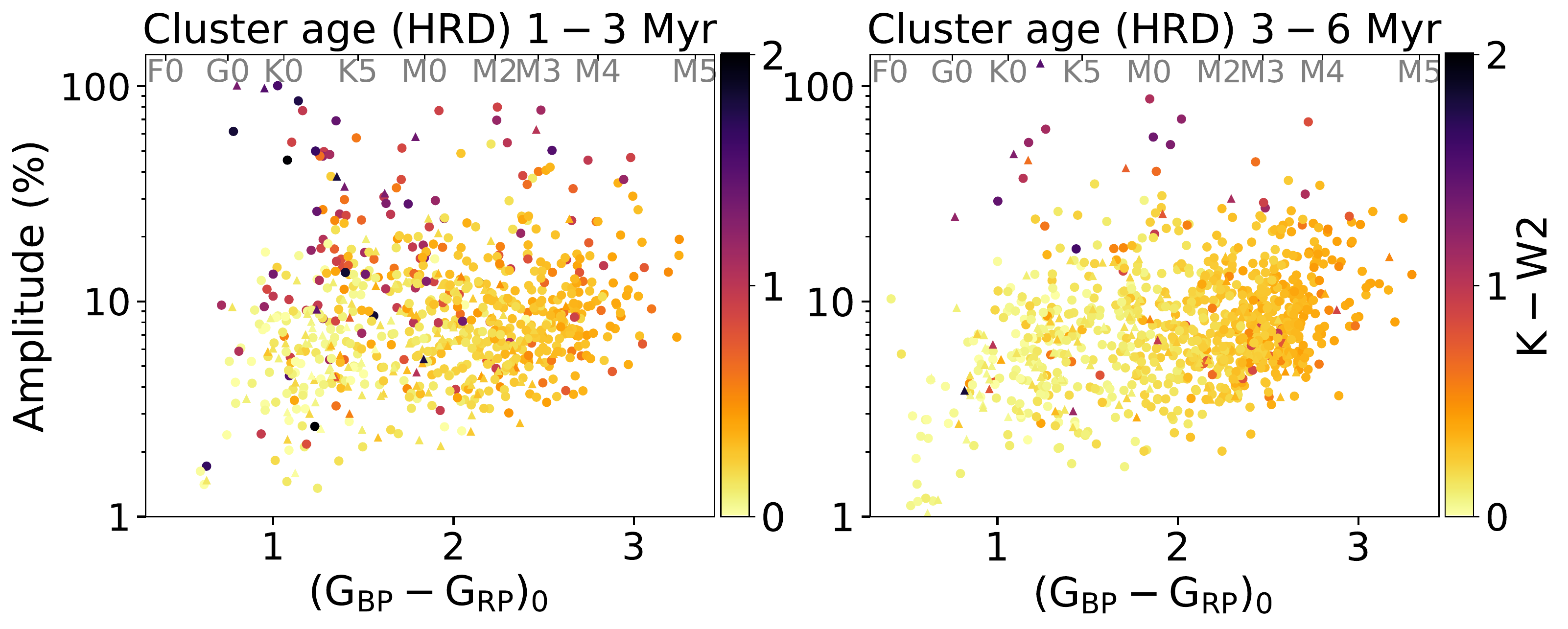}
    \caption{Amplitude vs $(G_{\mathrm{BP}}-G_{\mathrm{RP}})_{0}$ colour. Amplitude is calculated as 90th$-$10th percentile of the stellar flux, converted here to a percentage relative to the median flux level. Points are coloured by their $K-W2$ colour.}
    \label{fig:amplitude}
\end{figure}
\subsection{Amplitude}
In Figure \ref{fig:amplitude}, we plot amplitude (90th$-$10th percentiles of the flux, converted to a percentage) as a function of colour for the same selection as Figure \ref{fig:period_colour_density}. We see the smallest amplitudes appearing among the bluest stars, which may reflect the smaller spot coverage expected to be present, but we see some high-amplitudes present as well. These high-amplitude signals do, however, correspond with large values of $K-W2$ colour, indicative of a circumstellar disc, where the large flux variations likely originate from accretion bursters or dippers. We note that the increasing $K-W2$ colour trend from left to right is attributable to differing stellar photosphere shapes for different stellar masses, rather than being due to extinction by additional material in the system. Conversely, changes in $K-W2$ colour in the vertical direction, at a particular $(G_{\mathrm{BP}}-G_{\mathrm{RP}})_{0}$ colour, are indeed likely to be caused by material external to the photosphere. 

\subsection{Disc--rotation relation}
\label{section:discs}
\begin{figure}
	\includegraphics[width=0.9\columnwidth]{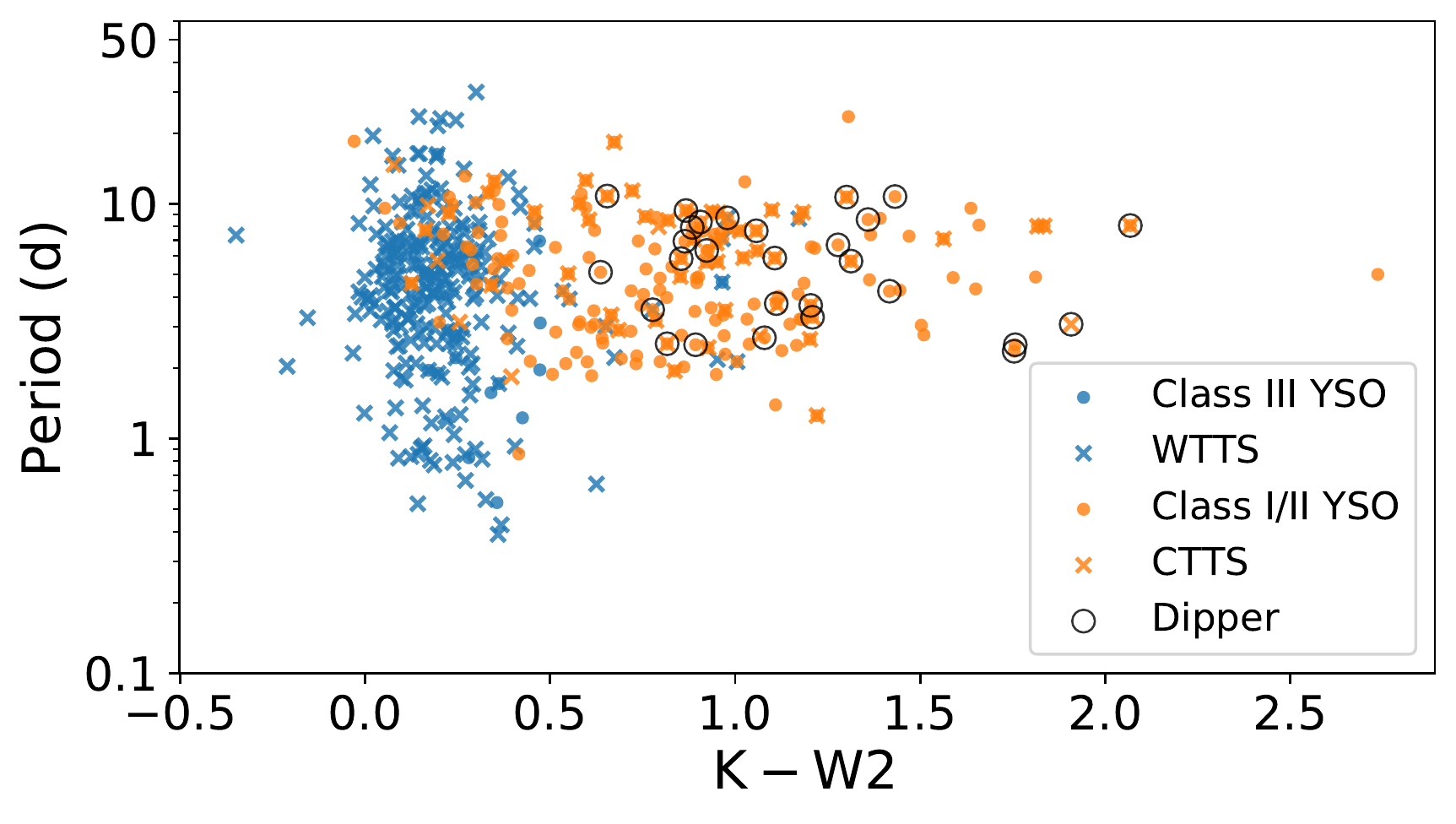}
    \caption{Rotation period vs $K-W2$ colour for objects with literature designations of either YSO class, T Tauri type, or dipper variable. Class III YSOs and WTTS are plotted with blue circles and crosses, respectively, while Class I or II YSOs and CTTS are plotted with orange circles and crosses. Identified dippers are shown with black open circles.}
    \label{fig:period_K_W2}
\end{figure}

Excess emission at infrared wavelengths -- thought to originate in the warm dust heated by irradiation from the central star -- is often used as an indicator for the presence of a circumstellar disc. In light of this, Figure \ref{fig:period_K_W2} shows the rotation periods as a function of $K-W2$ colour for stars with identifications found in the literature indicative of the presence or absence of a circumstellar disc. The blue markers locate objects classified as either Class III YSOs (blue circles) or WTTS (blue crosses), and the orange markers locate objects classified as either Class I or II YSOs (orange circles) or CTTS (orange crosses). Additionally, stars found to belong to the category of variable stars known as `dippers' -- objects displaying transient, aperiodic or quasi-periodic dimming events, possibly caused by a warped or clumpy inner-disc as seen from a nearly edge-on viewpoint \citep{Cody2014} --  are highlighted with open black circles \citep{Moulton2023}. 

The YSO classifications are based on photometry and were extracted from \citet{Hernandez2007}, \citet{Megeath2012} and \citet{Marton2016}. The T Tauri classifications on the other hand (sourced from \citet{Briceno2019} and \citet{Serna2021}) are derived spectroscopically, based on the relation between the equivalent widths of the H$\alpha$ line and spectral types. CTTS and WTTS labels distinguish stars which show or lack evidence of active accretion, respectively. While some WTTS may retain a passive, non-accreting circumstellar disc, the expectation is that there is a high degree of correlation between accretion and the presence of an inner circumstellar disc, as indicated by the Class I or II YSO designation, e.g. \citet{Nguyen2009} find accretion signatures based on H$\alpha$ equivalent widths to be highly correlated with 8 $\mu$m excess in 63/67 cases in their study of T Tauri stars in the young ($\sim$2 Myr old) Chamaeleon I and Taurus-Auriga star-forming regions. 

In Figure \ref{fig:period_K_W2}, we observe that the subset of stars displaced to the right, i.e. the population with significant infrared excess, is made up almost entirely of objects thought to be surrounded by a disc, rotating with periods longer than 2 d. To be precise, 4 per cent of CTTS or Class I/II YSOs rotate with periods shorter than 2 d, compared with 17 per cent for WTTS or Class III YSOs. In fact, the distribution of CTTS and Class I/II YSOs might more accurately be split at a point slightly below 2 d. Doing so at 1.8 d leaves just 1.5 per cent of CTTS or Class I/II YSOs below the cut, compared with 14 per cent of WTTS or Class III YSOs. The paucity of short-period rotators with significant infrared excess is consistent with the idea that disc braking plays an important role in the evolution of angular momentum in YSOs.

\section{Conclusions}
We conducted a $\sim$200-d monitoring campaign across 30 square degrees of the Orion Star-forming Complex. We determined probable members using astrometry from \gaia\ and corrected for extinction on a star-by-star basis. We reported periodicity 2268 out of 5749 stars and analysed rotation period distributions for 1789 stars with spectral types F0--M5. We assigned stars to clusters within Orion and determined their ages using MIST v.1.2 and Feiden magnetic evolutionary models. 

The vast majority of rotation periods lie in the range 1--10 d. We observe some evolution in period--colour space between younger and older populations. For older (3--6 Myr) clusters, we notice a shift towards shorter rotation periods for low-mass (>M2) stars, with no periods longer than 10 d among stars later than M3.5. This could indicate a mass-dependence in the dispersal of circumstellar discs. The turnover of the period--colour distribution also occurs at lower mass for the older-aged ensemble, e.g. we see the slow (90th percentile) rotators ($P_{\mathrm{rot}}\approx 10$ d) shift from $\sim$K5 (1--3 Myr) to $\sim$M1 spectral type (3--6 Myr). The fraction of binary candidates decreases towards redder colours in both young and old populations. 

Finally, we find that only 4 per cent (1.5 per cent) of CTTS and Class I/II YSOs rotate with periods shorter than 2 d (1.8 d), compared with 17 per cent (14 per cent) for WTTS and Class III YSOs.

\section*{ACKNOWLEDGEMENTS}
The NGTS facility is funded by a consortium of institutes consisting of 
the University of Warwick,
the University of Leicester,
Queen's University Belfast,
the University of Geneva,
the Deutsches Zentrum f\" ur Luft- und Raumfahrt e.V. (DLR; under the `Gro\ss investition GI-NGTS'),
the University of Cambridge, together with the UK Science and Technology Facilities Council (STFC; project reference ST/M001962/1). 

GDS gratefully acknowledges support by an STFC-funded PhD studentship.
EG gratefully acknowledges support from the David and Claudia Harding Foundation in the form of a Winton Exoplanet Fellowship, and from the UK Science and Technology Facilities Council (STFC; project reference ST/W001047/1).
JSJ greatfully acknowledges support by FONDECYT grant 1201371 and from the ANID BASAL project FB210003.

\section*{Data Availability}
The data underlying this article will be shared on reasonable request to the corresponding author.



\bibliographystyle{mnras}
\bibliography{paper} 




\appendix
\section{Injection--recovery test details}
\label{sec:appendix_inject}
The light curves for the injection--recovery tests were chosen to be from objects which had returned the systematic 1-day signal (or its aliases) from the period detection pipeline, i.e. stars apparently without a strong periodic signal of astrophysical origin. In order to ensure coverage across the full magnitude range, the original target stars for each field were supplemented with objects not in the members lists, but which had \gaia\ parallaxes placing them within the distance bounds of the cluster members. Outliers in plots of shot-noise vs magnitude for each field were then removed, leaving four sets of injection--recovery stars (one for each NGTS observation field).

For each star in the sample, the goal was to find the minimum amplitude of injected signal required for successful recovery. Doing this for a range of injected periods, would produce (recovered) amplitude distributions as a function of magnitude and period. The tests were conducted as follows. 35 periods were selected spanning 0.05 d to half the baseline of the observations, with a small random jitter added to each. 12 evenly-spaced samples were taken from phase space, again with random jitter. Then, for each star, for each period, for each phase, sinusoidal signals of increasing amplitude were injected, until the injected period was recovered in the Lomb-Scargle detection pipeline. The criterion for detection was that the recovered period fell within bounds based on the injected period, the baseline of observations and the sampling of the Lomb-Scargle algorithm:
\begin{equation}
\begin{split}
\mathrm{Bounds} = \Bigl\{\mathrm{min}\bigl({P_{\mathrm{inject}}-2\frac{P_{\mathrm{inject}}}{\mathrm{baseline}},\, P_{\mathrm{inject}}-3 dp}\bigr),\,\\ \mathrm{max}\bigl({P_{\mathrm{inject}}+ 2\frac{P_{\mathrm{inject}}}{\mathrm{baseline}},\, P_{\mathrm{inject}}+3 dp}\bigr)\Bigr\},
\end{split}
\end{equation}
where $dp$ is the Lomb-Scargle sampling in period space around the injected period.

\begin{figure}
	\includegraphics[width=\columnwidth]{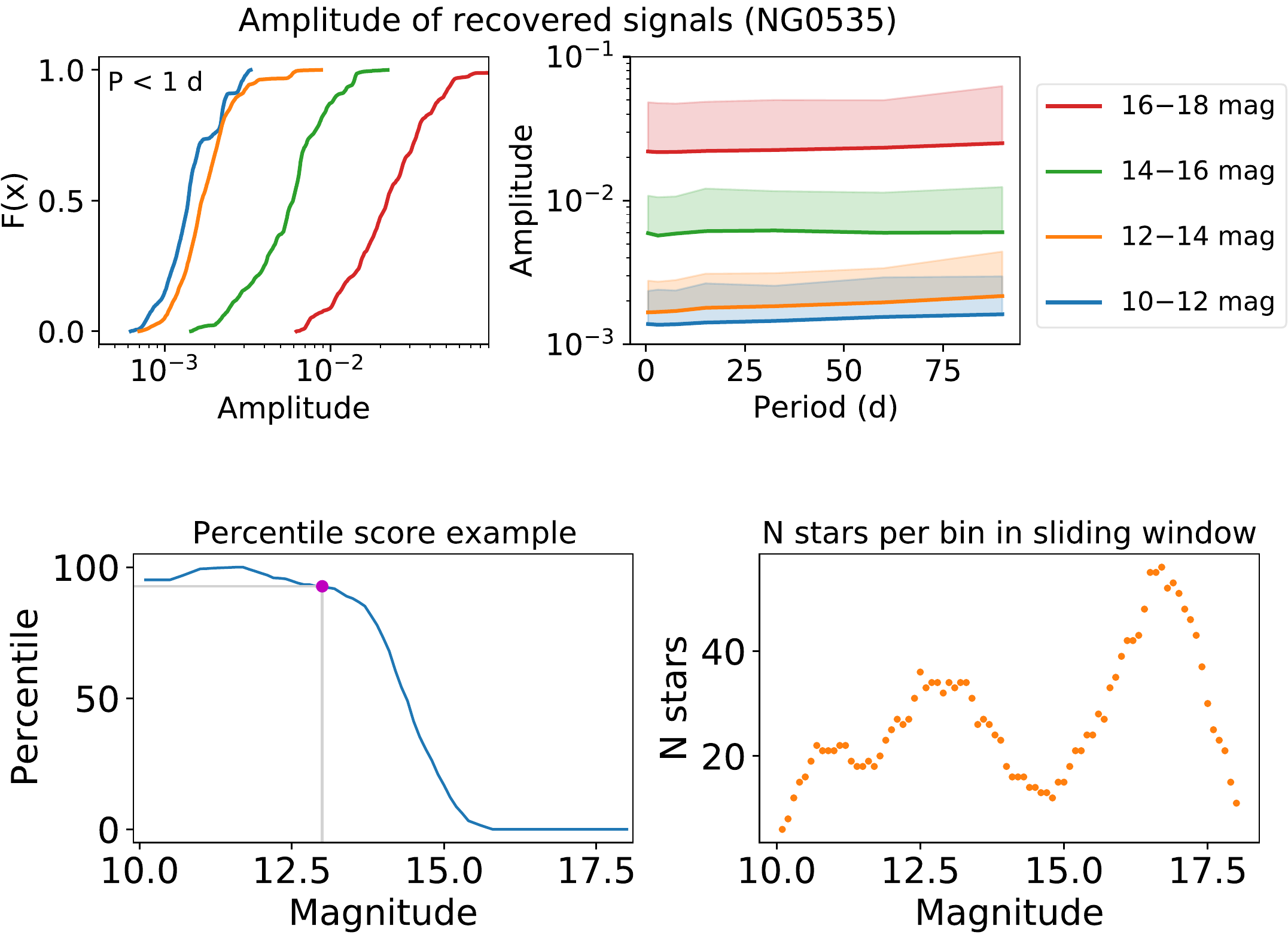}
    \caption{Top row: Two views on the injection--recovery results for field NG0535. Upper left: cumulative distribution functions for the amplitudes of successfully recovered signals with periods < 1 d. The colours correspond to different stellar magnitudes: blue ($G=$\,10--12), orange ($G=$\,12--14), green ($G=$\,14--16) and red ($G=$\,16--18) Upper right: Amplitude vs period for successfully recovered signals. Lines show the median values of the amplitude distributions and the shaded regions encompass 50--90th percentiles. Colour coding as before. Lower left: Percentile function example for a hypothetical detection of amplitude 0.003 and period of 5 d. The magenta point shows a star of magnitude 13, which would be given a score of 93. Lower right: Number of stars per bin in the sliding window. } 
    \label{fig:injection_recovery}
\end{figure}

Figure \ref{fig:injection_recovery} (top row) displays two views on the injection--recovery results for field NG0535 at a coarse level. The left-hand plot shows the cumulative distribution function for the recovered signal amplitudes, grouped into bins of size two stellar magnitudes, while the right-hand plot shows detected amplitude as a function of injected period. The most important variable is stellar magnitude. The Lomb-Scargle periodogram can be thought of in terms of least-squares fits around a constant reference model and a periodic model at each frequency, with best-fit sums of residuals $\chi^{2}_{\mathrm{ref}}$ and $\chi^{2}_{f}$, i.e.
\begin{equation}
P(f) \propto \chi^{2}_{\mathrm{ref}} - \chi^{2}_{f}.
\end{equation}
Hence, the periodogram peak height relative to the background noise depends primarily on the signal-to-noise of the data (i.e. the stellar magnitude) and the number of data points \citep{VanderPlas2018}.

The final stage of the process was to compare each periodic detection in the main sample to the injection--recovery results. For each object in the main sample, the detected amplitude of its best-fitting sinusoidal signal was compared with the amplitudes recovered in the injection--recovery tests, for stars of similar magnitude and for signals of similar period. The percentile of the detected amplitude relative to the injection--recovery amplitudes (at the corresponding magnitude and period) was then recorded as a score. The implementation was as follows. For each detection, take the injection--recovery results corresponding to the five nearest periods. Then, using a sliding window across the magnitude range of the injection--recovery sample, record the percentile of the target amplitude among the injection--recovery amplitudes for each step of the sliding window. The recorded percentile values were smoothed using a rolling mean filter, before the final percentile score for the target magnitude was obtained via linear interpolation. Figure \ref{fig:injection_recovery} (bottom row) shows an example of the results of a sliding window calculation for a particular amplitude and period.

\section{Effective temperatures: sourcing, corrections and uncertainties}
\label{sec:appendix_teff}
\subsection{Spectral type temperatures}
Spectral types from the literature were collected from \citet{Hillenbrand1997}, \citet{Sacco2008}, \citet{Hillenbrand2013}, \citet{Hsu2013}, \citet{Hernandez2014}, \citet{Skiff2014}, \citet{Koenig2015}, \citet{Fang2017}, \citet{Kounkel2017}, \citet{Briceno2019}, \citet{Manzo2020}, and \citet{Fang2021}. These were converted to $T_{\mathrm{eff}}$ by linear interpolation using the SC table described in section \ref{sec:stellar_params}. Spectral types were converted to integers for this process, i.e. 0--59 for classes B, A, F, G, K, M and their 10 subclasses. Where more than one spectral type was available for a source, the mean was used, avoiding duplicate values from compilation catalogues. Accompanying uncertainties in the spectral types were taken to be two subclasses for stars earlier than M0 and one subclass for M0 and later, which approximates the reported errors for YSOs in the Young Stellar Object Corral (YSOC; \citealt{YSOC}) (see Figure 8 in \citealt{Cao2022}). 

\subsection{APOGEE Net temperatures}
APOGEE Net is a deep convolutional neural network designed to predict $T_{\mathrm{eff}}$, log $g$ and Fe/H for stars with APOGEE spectra \citep{Apogee2018}. Version 1, described in \citet{Olney2020}, built on the data-driven approach of \citet{Ting2019}, which was trained on Kurucz atmospheric models, to incorporate training labels for PMS and low-mass MS stars based on empirical photometric relations and theoretical isochrones. It yielded properties for stars with $T_{\mathrm{eff}}<6700$ K in the DR14 APOGEE data release. \citet{Sprague2022} extended APOGEE Net to create a pipeline for estimating the parameters of stars across the full mass range in a self-consistent manner, applying it to DR17.

\begin{figure}
	\includegraphics[width=\columnwidth]{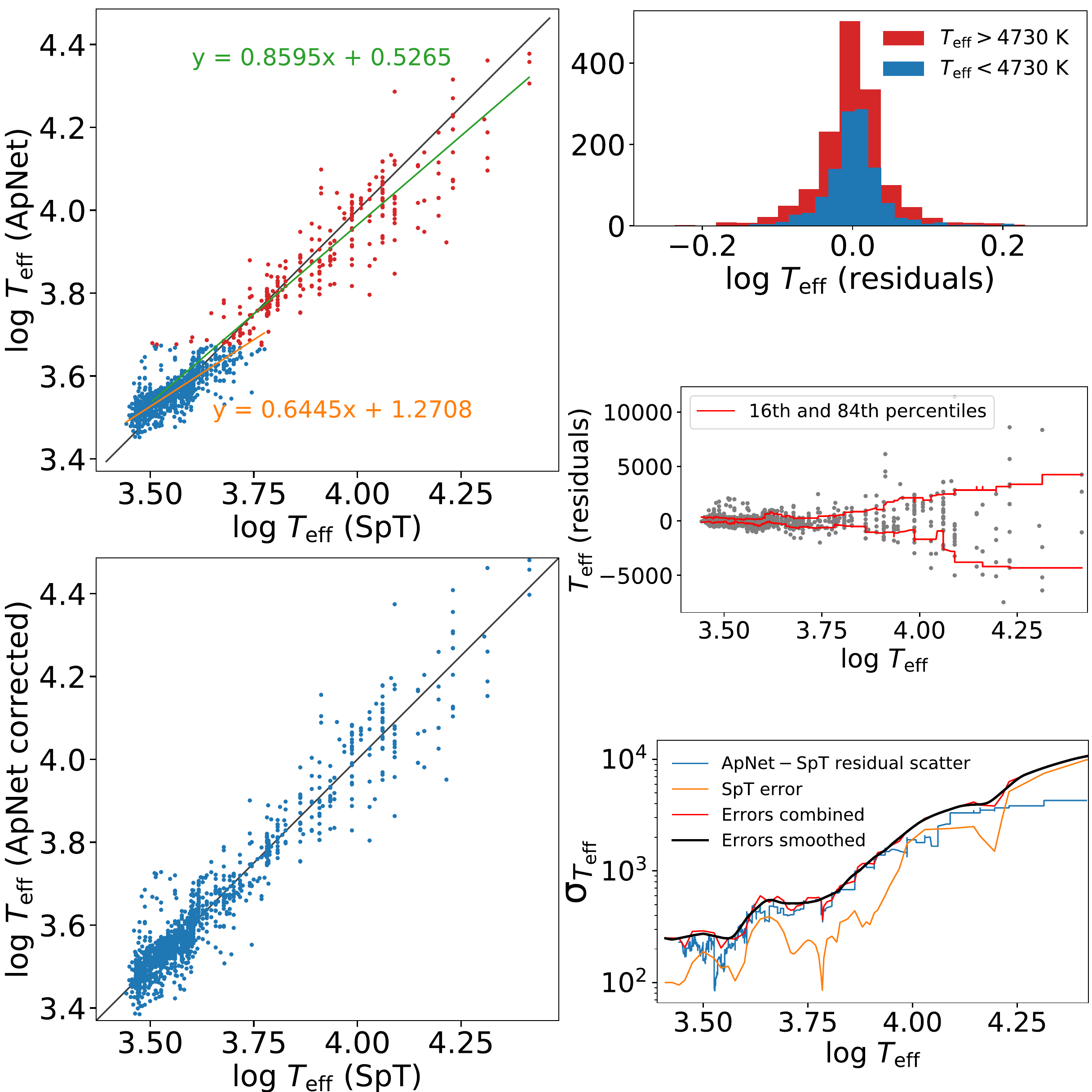}
    \caption{Top left: APOGEE Net log $T_{\mathrm{eff}}$ \citep{Sprague2022} plotted against log $T_{\mathrm{eff}}$ derived from literature spectral types, with a division at $T_{\mathrm{eff,ApNet}}=4730~\mathrm{K}$. Orthogonal distance regression lines for the two regions are over-plotted with their respective equations. Bottom left: log $T_{\mathrm{eff,ApNet}}$ vs log $T_{\mathrm{eff,SpT}}$ post correction. Top right: log $T_{\mathrm{eff,ApNet}}$ residuals (log $T_{\mathrm{eff,ApNet}}-T_{\mathrm{eff,SpT}}$). Centre right: $T_{\mathrm{eff}}$ residuals as a function of log $T_{\mathrm{eff}}$ along with rolling 16th and 84th percentiles. Bottom right: Combining (in quadrature) the scatter in the log $T_{\mathrm{eff,ApNet}}-T_{\mathrm{eff,SpT}}$ residuals with the spectral type errors to give $T_{\mathrm{eff}}$ error as a function of log $T_{\mathrm{eff}}$, used as a constraint in the MCMC (section \ref{sub:mcmc}).}
    \label{fig:apnet_sptype_corrections}
\end{figure}

In their study of $\lambdaup$ Orionis, \citet{Cao2022} noticed a trend in temperature in the cross-sample of sources with spectral types from the literature and APOGEE Net stars, which they attributed to the fact that the APOGEE Net PMS temperatures are generated from synthetic stars drawn from PARSEC isochrones. In our cross-sample of sources with both APOGEE Net and spectral type temperatures \footnote{To increase the cross-sample size, we use all sources within the cluster parallax bounds and precision previously described, i.e. not all objects in the cross-sample are in the cluster members list.}, which covers a much wider range of temperatures than the \citet{Cao2022} sample, we too find disagreement in $T_{\mathrm{eff}}$ between the two sources, but by way of two separate trends for low and high mass stars. The top-left plot in Figure \ref{fig:apnet_sptype_corrections} displays two linear fits in logarithmic temperature space using orthogonal distance regression, with the division being set at $T_{\mathrm{eff,ApNet}}=4730~\mathrm{K}$. The lower-left plot shows the cross-sample after correcting for these trends. So as to bring APOGEE Net temperatures in line with temperatures derived from spectral types, this correction was applied to all APOGEE Net stars used. Where both an APOGEE Net and a spectral type temperature existed, the spectral type temperature was adopted. The reason for the greater divergence from a 1:1 relation in the low-temperature domain is uncertain, but one potential contributing factor is the use, by APOGEE Net, of training labels made from synthetic photometry reliant on PARSEC v1.2S stellar models \citep{Chen2014}. PARSEC v1.2S models included a shift in the temperature--Rosseland mean optical depth relation, $T-\tau$, in order to reproduce the observed mass–radius relation for low-mass dwarf stars. However, such a correction may not simultaneously be a good recipe for contracting PMS stars in Orion. The shift was applied from 4730 K, increasing towards lower temperatures, which is our reason for placing the division at $T_{\mathrm{eff,ApNet}}=4730~\mathrm{K}$. Also in Figure \ref{fig:apnet_sptype_corrections}, is an illustration of how the residual scatter in the $T_{\mathrm{eff,ApNet}}-T_{\mathrm{eff,SpT}}$ relation was combined with the spectral type errors previously described. The residuals were fit with rolling 16th and 84th percentile filters across log $T_{\mathrm{eff}}$ space, with the mean of the (absolute) percentile values being added in quadrature with the spectral type errors, and then smoothed, to yield $\sigma_{T_{\mathrm{eff}}}$ as a function of $T_{\mathrm{eff}}$. The $\sigma_{T_{\mathrm{eff}}}$ values were then used as constraints on $T_{\mathrm{eff}}$ in the MCMC (section \ref{sub:mcmc}) and were applied to both APOGEE Net and spectral type temperatures. That is, Gaussian priors were placed on $T_{\mathrm{eff}}$, with mean values set to the spectral type or (corrected) APOGEE Net temperatures and standard deviations equal to $\sigma_{T_{\mathrm{eff}}}$.

\begin{figure}
	\includegraphics[width=\columnwidth]{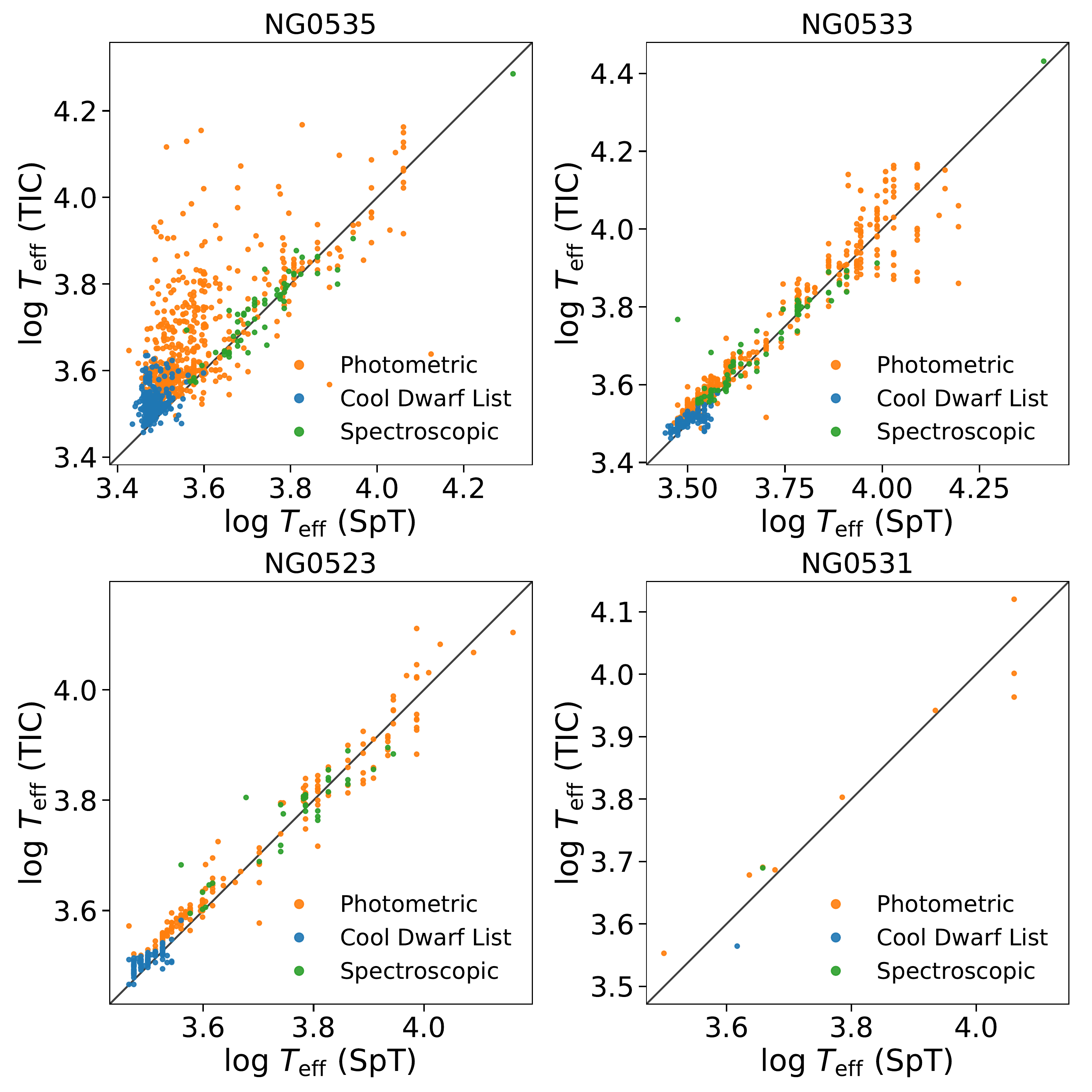}
    \caption{A comparison of effective temperatures from the TIC-8 catalogue with those obtained from literature spectral types for each of the four NGTS fields observed. Markers are coloured by the method by which each $T_{\mathrm{eff}}$ value was assigned in the TIC-8 catalogue: spectroscopic temperatures in green, temperatures from the Cool Dwarf List in blue, and temperatures from photometric relations in orange.}
    \label{fig:TIC_teff_refs}
\end{figure}

\begin{figure}
	\includegraphics[width=\columnwidth]{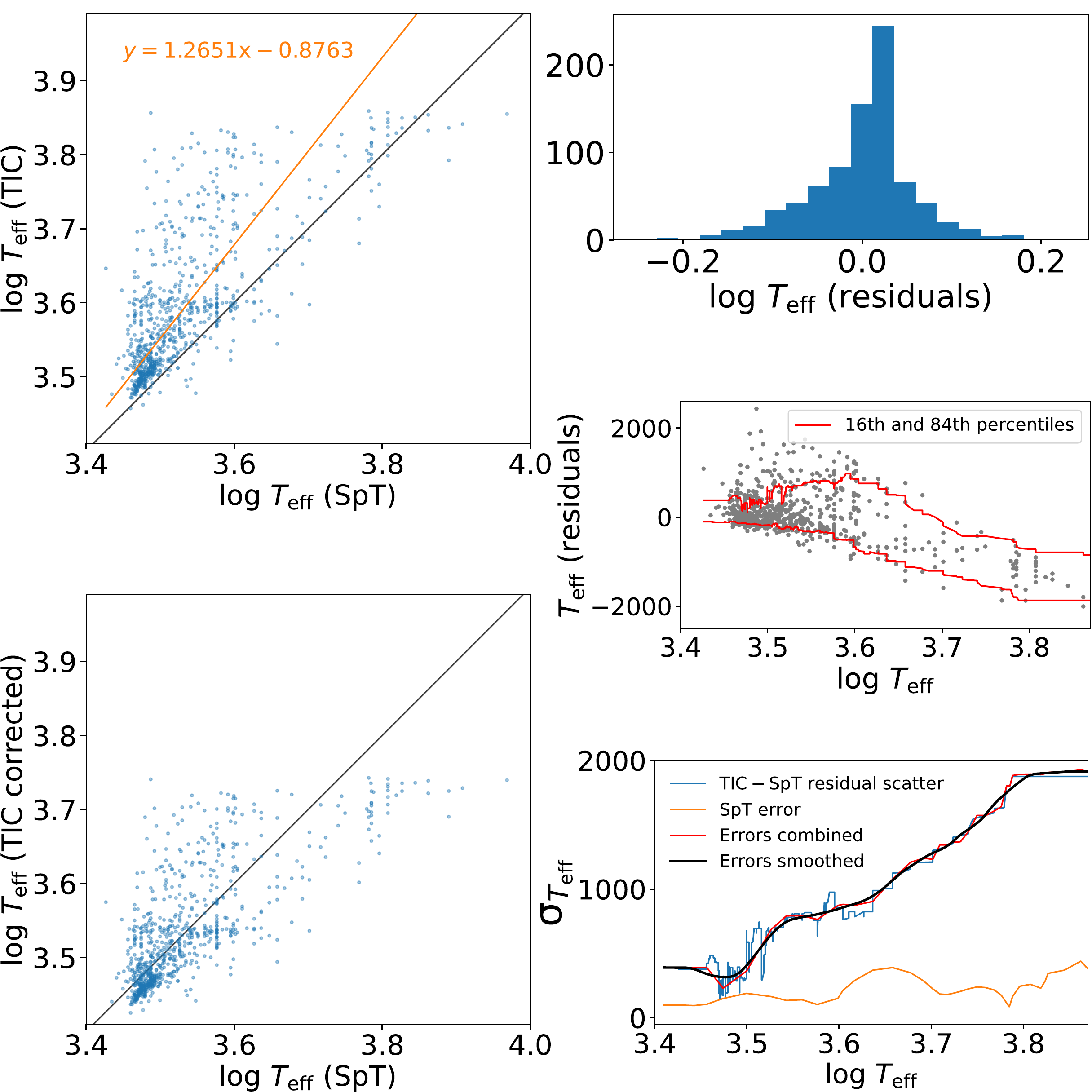}
    \caption{Equivalent to Figure \ref{fig:apnet_sptype_corrections}, but here comparing TIC-8 temperatures (excluding spectroscopic) below 7280 K in the NG0535 (ONC-centred) field with those from literature spectral types. Top left: TIC-8 log $T_{\mathrm{eff}}$ plotted against log $T_{\mathrm{eff}}$ from literature spectral types. Orthogonal distance regression line and equation overplotted. Bottom left: log $T_{\mathrm{eff,TIC}}$ vs log $T_{\mathrm{eff,SpT}}$ post correction. Top right: log $T_{\mathrm{eff,TIC}}$ residuals (log $T_{\mathrm{eff,TIC}}-T_{\mathrm{eff,SpT}}$). Centre right: $T_{\mathrm{eff}}$ residuals as a function of log $T_{\mathrm{eff}}$ along with rolling 16th and 84th percentiles. We take the maximum of the absolute values of the 16th and 84th percentiles of the plotted residuals, rather than the mean, as the error contribution here. Bottom right: Combining (in quadrature) the scatter in the log $T_{\mathrm{eff,TIC}}-T_{\mathrm{eff,SpT}}$ residuals with the spectral type errors to give $T_{\mathrm{eff}}$ error as a function of log $T_{\mathrm{eff}}$, used as a constraint in the MCMC (section \ref{sub:mcmc}).}
    \label{fig:tic_sptype_corrections}
\end{figure}

\subsection{TIC-8 temperatures}
In order to fit the stars without a sourced spectral type temperature or APOGEE Net temperature, we used the values from the TIC-8 catalogue. Effective temperatures in the TIC-8 catalogue are derived in three different ways for the sources in this work: from external spectroscopic catalogues, from the Cool Dwarf List (a carefully vetted list of stellar parameters for K- and M-dwarf stars with $T_{\mathrm{eff}}<4000$ K), or from photometric colours via empirical relations and a dereddening procedure (see \citet{Stassun2019} for details). Figure \ref{fig:TIC_teff_refs} shows how these temperatures compare with the available cross-sample of spectral type temperatures sourced from the literature. What is clear is that, whilst an approximate 1:1 relation is apparent in three out of the four fields, there is considerable disagreement for the ONC-centred field, NG0535, which is likely attributable to the high levels of extinction affecting observations of these stars. Because of this, we opted to treat TIC-8 temperatures for all fields except NG0535 in the same way as spectral type temperatures from the literature, with errors calculated as described above and displayed in Figure \ref{fig:apnet_sptype_corrections}. The same approach was also applied to TIC-8 spectroscopic temperatures for objects in field NG0535, i.e. to objects with spectroscopic temperatures in the TIC-8 catalogue, but where a spectral type temperature from the literature had not been sourced and no APOGEE Net temperature existed. For the other TIC-8 stars in field NG0535, with $T_{\mathrm{eff}}$ values from the Cool Dwarf List or the standard TIC-8 photometric relations, we attempted an equivalent procedure to that which was applied to the APOGEE Net temperatures (i.e. a linear correction), the results of which appear in Figure \ref{fig:tic_sptype_corrections}. We limited the correction to stars with TIC-8 temperatures below 7280 K ($\sim$F0 spectral type \citet{Pecaut2013}), where the scatter is reduced, and above which rotation by the detection of spot-modulation patterns in light curves is not expected. Two extreme outliers were also removed prior to the fit. Large amounts of scatter remain post correction, which is reflected in the final $T_{\mathrm{eff}}$ error estimates used as constraints in the MCMC and displayed in the bottom-right plot of the figure.



\bsp	
\label{lastpage}
\end{document}